\documentclass[aps,pre,superscriptaddress,twocolumn,preprintnumbers]{revtex4}
\usepackage[pdftex]{graphicx}
\usepackage{color}
\usepackage[bookmarks,bookmarksopen,pdfstartview=FitH]{hyperref}
\usepackage{amsmath}

\hypersetup{
colorlinks=true,
urlcolor=blue, linkcolor=red, citecolor=blue}    
\usepackage{comment}

\begin{document}

\author{Paul Z. Hanakata\footnote{current address: Department of Physics, Boston University, Boston, MA 02115 }}
\affiliation{Department of Physics, Wesleyan University, Middletown, CT 06459, USA}
\author{Beatriz A. Pazmi\~no Betancourt}
\affiliation{Department of Physics, Wesleyan University, Middletown, CT 06459, USA}
\affiliation{Materials Science and Engineering, National Institute of Standards and
Technology, Gaithersburg, Maryland 20899, USA}
\email{bpazminobeta@wesleyan.edu}
\author{Jack F. Douglas\footnote{Official contribution of the U.S. National Institute of Standards and
  Technology - Not subject to copyright in the United States}}
\affiliation{Materials Science and Engineering, National Institute of Standards and
Technology, Gaithersburg, Maryland 20899, USA}
\author{Francis W. Starr}
\affiliation{Department of Physics, Wesleyan University, Middletown, CT 06459, USA}

\title{A Unifying Framework to Quantify the Effects of Substrate Interactions, Stiffness, and Roughness on the
  Dynamics of Thin Supported Polymer Films } 

\date{\today}

\begin{abstract}
  Changes in the dynamics of supported polymer films in comparison to
  bulk materials involve a complex convolution of effects, such as
   substrate interactions,  roughness and compliance,
  in addition to film thickness. We consider molecular dynamics
  simulations of substrate-supported, coarse-grained polymer films where
  these parameters are tuned separately to determine how each of these variables
  influence the molecular dynamics of thin polymer films. We find that all these variables
  significantly influence the film dynamics, leading to a seemingly
  intractable degree of complexity in describing these changes.  However, by
  considering how these constraining variables influence string-like
  collective motion within the film, we show that all our observations
  can be understood in a unified and quantitative way.  More specifically,
  the string model for glass-forming liquids implies that the changes in the structural relaxation of these
  films are governed by the changes in the average length of string-like cooperative  motions and this model is confirmed under all conditions considered in our simulations.
 Ultimately, these changes are parameterized in terms of just the activation enthalpy and entropy for molecular organization, which have predictable dependences on substrate properties and film thickness, offering a 
 promising approach for the rational design of film properties.
\end{abstract}
\maketitle
\section{Introduction}
Polymer films are used in a wide variety of applications, ranging from
micro-electronic devices to artificial tissues~\cite{litho-rev06,
  tissue-eng}.  However, both mechanical and dynamical properties of
polymeric materials often change considerably in relation to bulk
once confinement dimensions become less than $\simeq100$ nm.  Much of
the effort aimed at understanding the property changes in thin
polymer films has centered on measurements related to the stiffness of
these films~\cite{stafford06, connell08} and changes of molecular
mobility, as quantified by the glass-transition temperature
$T_g$~\cite{mckenna-rev05, confinement-rev}.  Many
experimental~\cite{keddie1, fdd97, fukao01, depablo01, et03,
  mckenna-rev05, kumar05, tork05, tork09}, as well as
computational~\cite{depablo00-2, smith03, basch05, rigg-prl06,
  basch09, basch-soft-rev10} studies, have reported large property
changes in thin films. These changes have been mainly attributed  to a
combination of substrate interaction and geometrical
confinement. There is also a growing awareness of the relevance of
substrate roughness and stiffness, as well as non-equilibrium
residual stress effects in cast films.  It is a difficult matter to
separate all of these different effects in experiments, and the
present work addresses this general problem through molecular dynamics
simulations of substrate-supported, coarse-grained polymer melt films
of variable thickness where the polymer-substrate interaction is
varied, along with the boundary roughness and rigidity.  Since we can
tune these parameters in simulations, we can obtain clear indications
of how each of these variables influence the film molecular
dynamics.  After an analysis of how these diverse factors affect basic
dynamic properties of the polymer film, we show that the dynamical
changes under all these conditions can be organized and understood in
terms of how these constraining variables influence
collective motion within the film, parameterized by the enthalpy and entropy of activation for molecular reorganization.

Changes in $T_g$  in polymer films are usually associated
with local changes in the dynamics near the interfaces. Many studies have reported
that a repulsive or neutral substrate  along with a free boundary leads to an
enhancement  in dynamics and  a reduction of $T_g$~\cite{forrest96, fdd97,
basch09,  paeng12}.   In  contrast,  an attractive  substrate,  which
typically  slows down  the  dynamics near  the  substrate, results  an
increase   in   $T_g$~\cite{depablo00-2,   depablo01,  et03,   tork05,
tork02-2,   basch-soft-rev10,   schon03,  mckenna-rev05,fdd97,   kj01,
kumar05,  basch05}.   However, an  attractive  smooth  substrate with  a
relatively weak  interaction may also  enhance the rate  of relaxation
and  diffusion~\cite{basch06,   depablo00-2,  tork05,  et03,  fukao01,
hanakata12},  demonstrating  that  the  polymer-substrate  substrate
strength and the substrate roughness can also have significant effects on the
polymer film dynamics.
In particular, it has been noted that the enhancement or slowing of
relaxation in supported films induced by two interfaces with different
properties can complicate the interpretation of the thickness
dependence of $T_g$~\cite{kj01, fdd97, tork02-2, basch-soft-rev10,
  basch09, forrest96, schon03, et03, hanakata12}.

  The most prevalent
type of polymer films are those supported on solid substrates, where
the relaxation time is often increased near the substrate, while decreased at
the free boundary.  Additionally, experiments on multi-layered interfacial films have
shown that the effects of the free boundary region can be largely eliminated
by placing films between stacks of nano-layered polymers with different
species~\cite{tork07}, suggesting that there is a length scale associated
with the interfacial film dynamics.  This leads to the question of whether
the film dynamics depends simply on the substrate interaction, or are
there other physically relevant characteristics of the interface that
must be considered. After all, glass formation is a dynamical
phenomenon, so that other variables -- such as
substrate rigidity -- might be relevant. This motivates an exploration of
the effects of substrate rigidity on properties of thin polymer films, a
property that can be greatly tuned in polymeric materials through
cross-linking or through control of the molecular
structure~\cite{tate01, erber10, tong13}

A popular picture to rationalize the changes of the film dynamics is a
superposition of polymer layers with locally varying dynamics. In this
simple model, any changes in the overall dynamics should be manifested
locally. Thus, the interfacial layers are correspondingly expected to
be the primary contributor to changes in the overall film
dynamics.  Near an attractive substrate, the polymers are `bound' to the
surface, leading to slower dynamics, while at the free boundary region of a
supported or free-standing film, the chains have a relatively higher
mobility.  At the film center, far from both interfaces, the local
properties are expected to be `bulk-like'. This layer picture of
film dynamics is often conceptually linked to local changes in density profile or free volume. In our previous work, we found inconsistencies for
this free-volume layer (FVL) rationale for explaining the observed
changes in the dynamics~\cite{hanakata12}. Moreover, the dynamics
can be enhanced at the substrate, despite an increase in local density. 
We also quantified the length scales of both density and
dynamical perturbations within supported films and found that the
temperature dependence of these scales are opposite to that at the
free boundary region, inconsistent with the  FVL approach. The
changes in the dynamics of the film with a supporting layer are
generally \emph{non--local}, so it is naive to treat the film interior as being
the same as bulk material.

Here, we  consider the dependence of the
dynamics on film thickness, substrate roughness, and rigidity.  We find that these parameters can induce significant
changes in the dynamics, characterized by changes in the film $T_g$
and fragility, but only rather subtle changes are observed in static
properties, such as density.  Again, we find that free volume
ideas are not  useful in predicting dynamics at the local
level. Rather,  substrate interaction, substrate
roughness, and stiffness can all greatly influence the mobility
gradient transverse to the substrate.  Our findings for the variation
of $T_g$ with substrate roughness and interaction strength are
consistent with earlier works~\cite{depablo00-2, depablo01, michels11,
  michels12}, but our observations on fragility and regarding
substrate stiffness are new. Another novel aspect of the
current work is that we characterize the fragility changes in
terms of cooperative motion within the film and, in this way, obtain a
\emph{quantitative} understanding of the wide variations in the temperature
dependence of the structural relaxation time with film boundary
conditions and thickness.

There is continued interest in the breakdown of the Stoke-Einstein relation
in glass-forming liquids and the possible relation of this phenomena with fragility and dynamical heterogeneity, and several recent studies have suggested specific relationships. Since we are able to tune the fragility over a large range using the same polymer model through modifications of confinement, we can asses the validity of these relations in our glass-forming polymer model.  We find that the decoupling exponent relating the structural relaxation time to a diffusion relaxation time can be systematically varied with confinement. The degree of decoupling increases as the effective dimension is reduced, i.e., smaller film thickness, consistent with recent observations for  model glass-forming liquids in a variable spatial dimension~\cite{sastry13}. Moreover, film fragility can either increase or decrease under confinement, depending on the boundary interaction, so we do not generally see an increase in decoupling with greater fragility, as suggested by crystallization measurements in non-polymeric materials \cite{ediger08Decoupling}. Our results support recent observations~\cite{sastry13} that indicate that changes in spatial dimensionality are relevant to understanding the decoupling phenomenon.

Given the sensitivity of the dynamics to the large collection of
substrate properties, the question remains: how do we obtain a unified
understanding of all these effects on the polymer dynamics? There has
been much speculation that these changes revolve around changes in the
collective dynamics of the polymer molecules, where the Adam-Gibbs
theory is often discussed without a specific definition of the hypothetical `cooperatively rearranging regions' (CRR)
that are relevant to understanding these property changes.
Simulations have identified cooperative rearrangements 
that are quantitatively linked  to the structural relaxation
time for bulk polymer materials~\cite{sds13}, and a similar connection has also been established in model
polymer nanocomposites~\cite{sd11, beatriz13}.  These string-like motions therefore offer a molecular realization of the abstract CRR. We test this
predictive scheme for the molecular dynamics simulations of
polymer films where the inherent inhomogeneity of the dynamics of
these materials makes it unclear whether is the model should still
apply. Encouragingly, we obtain a remarkable reduction of all our
simulation data for structural relaxation in thin polymer films based
on this unifying framework. Lastly, we investigate the influence
of confinement on the activation free energy parameters that define our description.

\section{Modeling and Simulation}
We model polymers  as unentangled chains of beads linked by harmonic springs.  The substrate is modeled either as a collection of substrate atoms, or by a perfectly smooth substrate.
Non-bonded monomers or atoms of the substrate interact with each other
via  the Lennard-Jones  (LJ) potential,  and  we use  a shifted-force
implementation to  ensure continuity of  the potential and  forces at
the cutoff distance $r_{\rm  c}$. We choose $r_{\rm c}=2.5\sigma_{ij}$
to  include inter-particle  attractions where  $\sigma_{ij}$ is  the monomer
``diameter'' in  the LJ potential.  The index  pair $ij$ distinguishes
interactions  between  monomer-monomer  (mm), substrate-monomer  (sm),  and
substrate-substrate (ss) particles.  The LJ  interaction is not included for the
nearest-neighbors along  the chain.  These monomers are  connected by a
harmonic    spring   potential    $U_{\rm    bond}   =    \frac{k_{\rm
chain}}{2}\big(r - r_{\rm 0} \big)^2$  with bond length $r_{\rm 0} =
0.9$  (equilibrium  distance) and  spring  constant  $k_{\rm chain}  =
(1111)\epsilon_{\rm   mm}/\sigma_{\rm mm}^2$.  $r_{\rm  0}$.   The  spring
constant  is chosen  as  in Ref.~\cite{basch06},  but we choose $r_{\rm 0}$ smaller
than in Ref.~\cite{basch06} because  we  found  crystallization  occurs
readily in the films for the value used in Ref.~\cite{basch06}.

The interaction between monomers and the smooth substrate is given by,
\begin{equation}
V_{\rm smooth} = \frac{2\pi}{3}\epsilon_{\rm sm}\rho_{\rm
  s}\sigma_{\rm ss}^3 \left[
  \frac{2}{15}\left(\frac{\sigma_{\rm sm}}{z}\right)^9 -
  \left(\frac{\sigma_{\rm sm}}{z}\right)^3 \right],
\label{eq:substrate-smooth}
\end{equation}
where $z$ is the distance of a monomer from the substrate. This is the same
smooth  substrate    model   that    we    studied    in   our    previous
work~\cite{hanakata12}.  To  model the rough substrate, we  tether the substrate
atoms  to the  sites of  triangular lattice  (the 111  face of  an FCC
lattice) with harmonic potential,
\begin{equation}
\\U_{\rm s}(r_i) = \frac{k_{\rm s}}{2}\bigg(|{\vec r_i} - {\vec r_{i\rm eq}}| \bigg)^2,
\label{U_{substrate}}
\end{equation} 
where ${\vec r_{\rm eq}}$ denotes an equilibrium position on the
triangular lattice and $k_{\rm s}$ is the harmonic spring
constant~\cite{basch05}.  We choose the lattice spacing to be
$2^{1/6}\sigma_{\rm ss}$, where $\sigma_{\rm ss} = 0.80\sigma_{\rm
  mm}$ and $\sigma_{\rm sm}=\sigma_{\rm mm}$.  All values are in reduced units, where $\sigma_{\rm mm}=1$ and
$\epsilon_{\rm mm}=1$.  Varying $k_{\rm s}$ allows us to examine the
role of substrate rigidity on the polymer dynamics.  We simulate films
of variable thicknesses with $N_c$ = 200, 300, 400, 600, or 1000
chains of 10 monomers each.  These sizes correspond to thicknesses
with value of roughly 6 to 25 monomer diameters.  We use various
 interaction strengths ($\epsilon_{\rm sm}\equiv\varepsilon$) between the rough
substrate and polymers, ranging from 0.4 to 1.0 $\epsilon_{\rm mm}$
with a fixed surface rigidity $k_s = 100$; we vary the strength of the substrate rigidity ($k_s\equiv k$) over the range from 10 to 100 with a fixed $\varepsilon = 1$. For this
range of model parameters, we find $T_g$ of the film can be higher or
lower than the bulk value.  Additionally, we simulate a pure bulk
system of 400 chains of $M = 10$ monomers each at zero pressure for the purpose of comparison.

We define film thickness $h(T)$ as a distance from the substrate where
the density  profile along the $z$ direction, perpendicular to the substrate, $\rho(z)$  decreases  to 0.10.   Other reasonable  criteria  does not
affect  our  qualitative  findings.    The  resulting  $h(T)$  is  well
described by an Arrhenius form,  which we use to extrapolate the thickness value
$h_g \equiv h(T_g)$ at the glass transition.

To  quantify the overall  dynamics of  the films  and bulk  system, we
evaluate the coherent intermediate scattering function,
\begin{equation}
F(q,t)\equiv\frac{1}{NS(q)}\Bigg\langle\sum_{j,k_s=1}^{N}e^{-iq.[r_k(t)-r_j(0)]}\Bigg\rangle
\label{eq:fqt}
\end{equation}
where $r_j$  is the position of  monomer $j$ and $S(q)$  is the static
structure  factor.   We  define  the  characteristic  time  $\tau$  by
$F(q_0,\tau) = 0.2$, where $q_0$ is  the location of the first peak in
of $S(q)$.  To  quantify dynamics locally within the  film, we use the
self (or incoherent) $F_{\rm{self}}(z,q,t)$  part ({\it i.e.\ } $j=k$)
of Eq.~(\ref{eq:fqt}) on the basis of the position $z$ of a monomer at
$t=0$.     We   define    the   relaxation    time    $\tau_{\rm s}(z)$   by
$F_{\rm{self}}(z,q_0,\tau_{\rm s}) = 0.2$.

\section{Dependence of $T_g$ and Fragility on substrate Structure}
\subsection{Survey of Substrate Roughness and Film Thickness Effects}
\begin{figure}[ht]
\centering\includegraphics[trim=0mm 15mm 0mm 15mm, clip,width=0.48\textwidth]{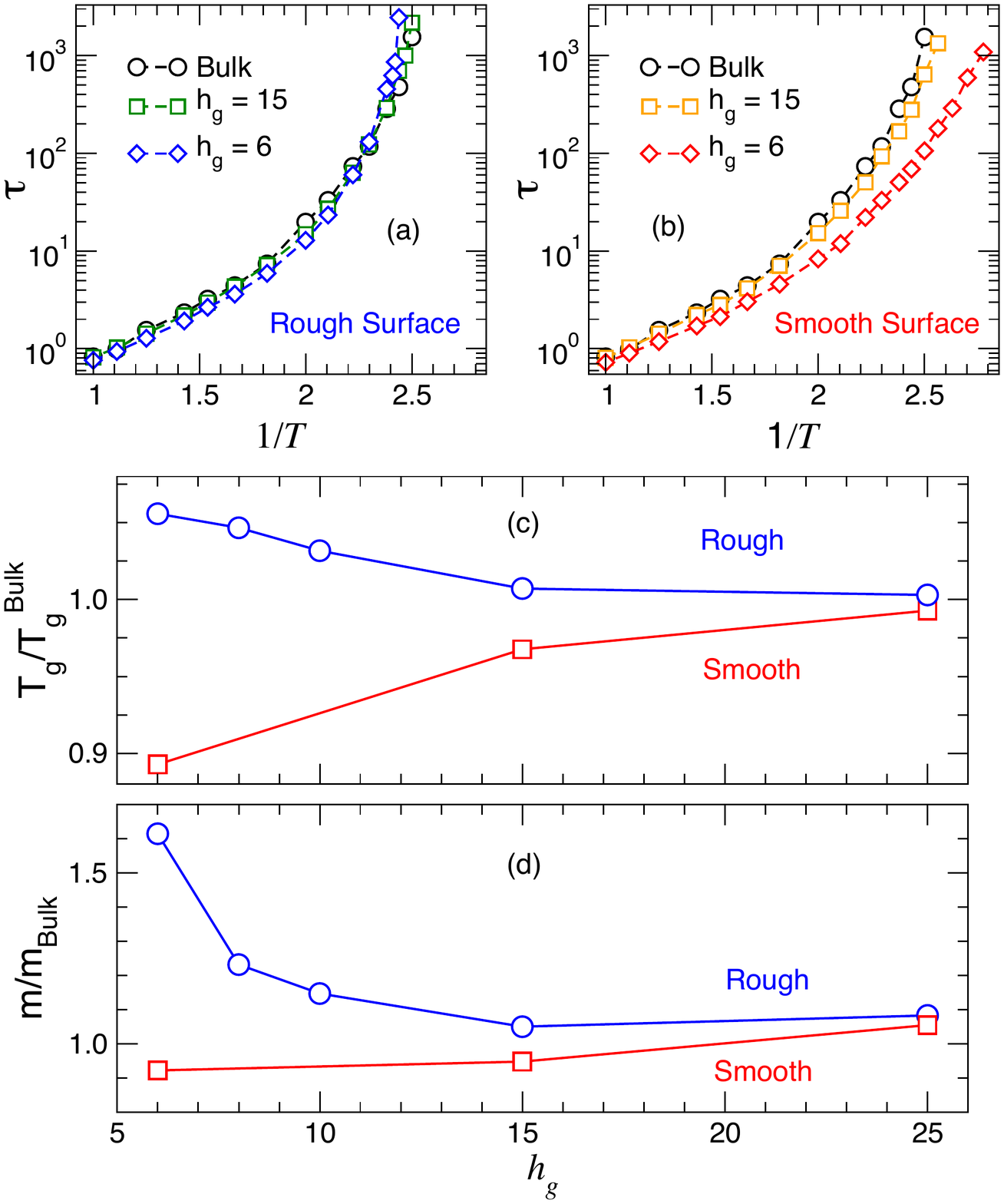}
\caption{Effects of film thickness and structure of the supporting
  substrate on glass transition temperature $T_g$ and fragility.  The
  $T$ dependence of relaxation time $\tau$ of a bulk system and two
  representative film thicknesses $h_g$ supported on (a) a rough or (b) smooth
  substrate ($\varepsilon =1$).  In this $T$ range, it is apparent that, relative to the
  bulk, $\tau$ increases as the film thickness is decreased on rough substrates, while for smooth substrates shows an opposite behavior. (c) Relative
  $T_g$ and (d) relative fragility $m$ to the bulk as function of film
  thickness.  Both $T_g$ and $m$ for the rough substrates
  increase, while $T_g$ and $m$ of smooth substrates decrease as we
  decrease film thickness.}
\label{fig:tau-tg-rs}
\end{figure}
We first contrast the overall changes to glass formation of polymer
films with various thicknesses supported on a rough or smooth
substrate having the same substrate-monomer interaction ($\varepsilon =
1.0 $).  Relative to the bulk system, the relaxation time $\tau$ of
polymer films on the smooth substrate decreases as we decrease film
thickness, and these deviations become more pronounced as we go to
lower $T$, consistent with previous studies~\cite{hanakata12, basch06}
(Fig.~\ref{fig:tau-tg-rs}(b)).  However, we find the opposite trend
 for the rough substrate, as noted in Ref.~\cite{scheidler02,
  virgiliis12}.  As we will see, this trend depends on substrate interaction strength and rigidity. We see that the dynamics change more rapidly with $T$
for thinner films resulting in a larger $\tau$ relative to the
bulk material (Fig.~\ref{fig:tau-tg-rs}(a)).  We estimate $T_g$ by fitting our data to the Vogel-Fulcher-Tammann (VFT)
equation,
\begin{equation}
\tau(T)=\tau_\infty e^{DT_0/(T-T_0)}.
\label{eq:VFT}
\end{equation}
where $\tau_\infty$ is an empirical prefactor normally on the order of a molecular vibrational time ($10^{-14}$ to $10^{-13}$s)~\cite{Angell97}, $D$ is a measure of `fragility' and $T_0$ is a temperature at which $\tau$ extrapolates to infinity. Eq.~\ref{eq:VFT} should only be applied above the glass transition temperature.
In  a  lab setting,  $T_g$  is  often defined  as  $T$  at which  the
relaxation  time  reaches $100$ s~\cite{angell95},  and  we adopt  this
simple  criteria.  Figure~\ref{fig:tau-tg-rs}(c)  shows  that,
relative  to the bulk,  $T_g$ of  polymer films  on the  rough substrate
increases with decreasing film thickness, while for the smooth substrate
systems, $T_g$ decreases with decreasing film thickness.

The  variation  in  $T$  dependence  of relaxation  is  quantified  by
fragility,  defined as  the logarithmic slope  of relaxation  time at
$T_g$
\begin{equation}
m(T_g)  \equiv \left. \frac{\partial \ln  \tau}{\partial (T/T_g)}
\right|_{T_g}.
\label{eq:m}
\end{equation}
We  evaluate fragility $m$  using the  fit of  Eq.~(\ref{eq:VFT}).  In
Fig.~\ref{fig:tau-tg-rs}(d),  we  see that,  relative  to the  bulk,
films on the rough  substrate become more fragile as  we decrease thickness,
which is  apparent from the increasingly rapid  variation of $\tau(T)$
(Fig.~\ref{fig:tau-tg-rs}(a)).   In   contrast,  the  fragility  of
polymer  films on the  smooth substrate  decreases weakly with  decreasing film
thickness.

Experimentally, $T_g$ is often found to be proportional to
$m$~\cite{mckenna}.  We also find a correlation between $T_g$ and $m$
for both substrates, but this relation is not strictly proportional.
Note that films supported on a smooth substrate may have a non-monotonic
thickness dependence of $T_g$ and $m$ on thickness.  Specifically, our
recent work~\cite{hanakata12} showed that $T_g$ or $m$ decreases with
decreasing film thickness on the smooth substrate up to some critical thickness, but 
that $T_g$ increases for very thin films when interfacial effects become
dominant.
\subsection{Local Structure and Dynamics}
\begin{figure}[ht]
\centering\includegraphics[trim=0mm 80mm 0mm 10mm, clip,width=0.48\textwidth]{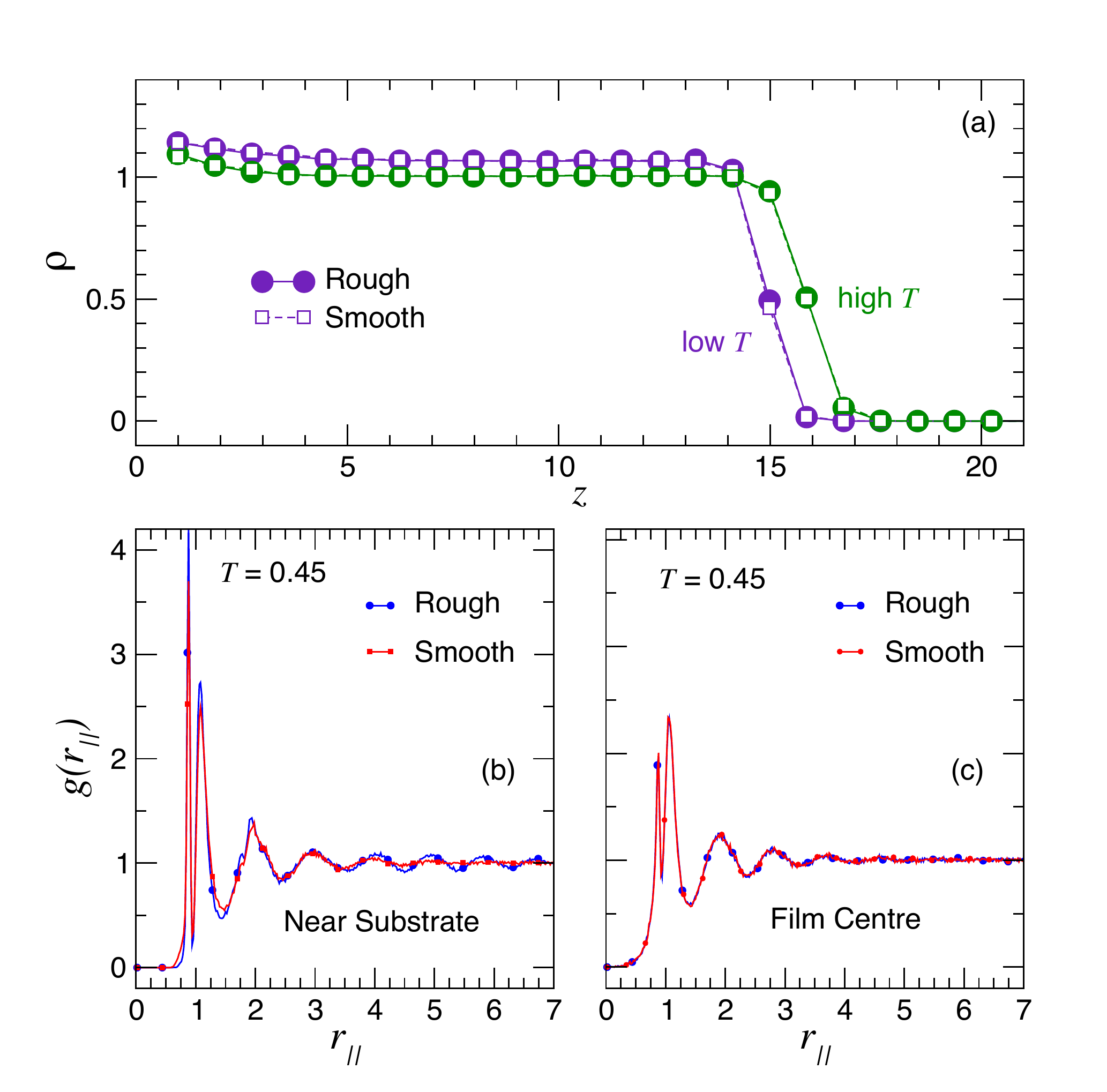}
\caption{(a) Monomer density profile $\rho(z)$ of a film, $h_g=15$,
  supported on a rough or smooth substrate. (b) Pair--pair correlation function in the direction parallel to the substrate
  $g(r_{||})$  near the substrate. (c) $g(r_{||})$  at the film center.   Monomers near the rough substrate are
  slightly more densely packed and have better local ordering in
  comparison to those near the smooth substrate.}
\label{fig:local-all-rs-1}
\end{figure}
To understand the observed  changes in $T_g$ and fragility, we
resolve both structure  and dynamics locally, since the  changes in the properties of the
film as a whole should be manifested in its local properties. We first
contrast  the  local  dynamics  and  monomer density  as  function  of
distance $z$ from  the substrate boundary of rough or smooth substrates
with  monomer-substrate interaction strength $\varepsilon=  1$ .   We  evaluate   both  $\rho(z)$  and
$\tau_{\rm s}(z)$ with a bin size $\delta z=0.875$.

In  figure~\ref{fig:local-all-rs-1}(a), we  observe that  the monomer
density near either the smooth or rough substrate increases  weakly, and has a steady value
through most of the film. The density drops to zero over a narrow
window at the free boundary region. At the center of the film, the density has
a value close  to the bulk. The density profile of  the film on smooth
substrate is essentially identical to that of a film on a rough substrate. 

In addition, we contrast the local structure parallel to the
substrate by evaluating the density pair correlation function
$g(r_{||})$ (see Fig.~\ref{fig:local-all-rs-1}(b) and (c)).  Far from
the substrate, $g(r_{||})$ of both systems is indistinguishable, as the
monomers are completely unperturbed by the substrate.  Near the
substrate, we see that there is a slight difference in the local
structure. In particular, near the substrate, $g(r_{||})$ of the rough substrate has a
somewhat larger first peak, indicating that the monomers near the rough substrate
are more ordered than those near the smooth substrate.  In
addition, there is a weak long-range ordering of monomers for the
rough substrate, potentially induced by the periodicity of the substrate
atoms. 

\begin{figure}[ht]
\centering\includegraphics[trim=0mm 160mm 0mm 0mm, clip,width=0.48\textwidth]{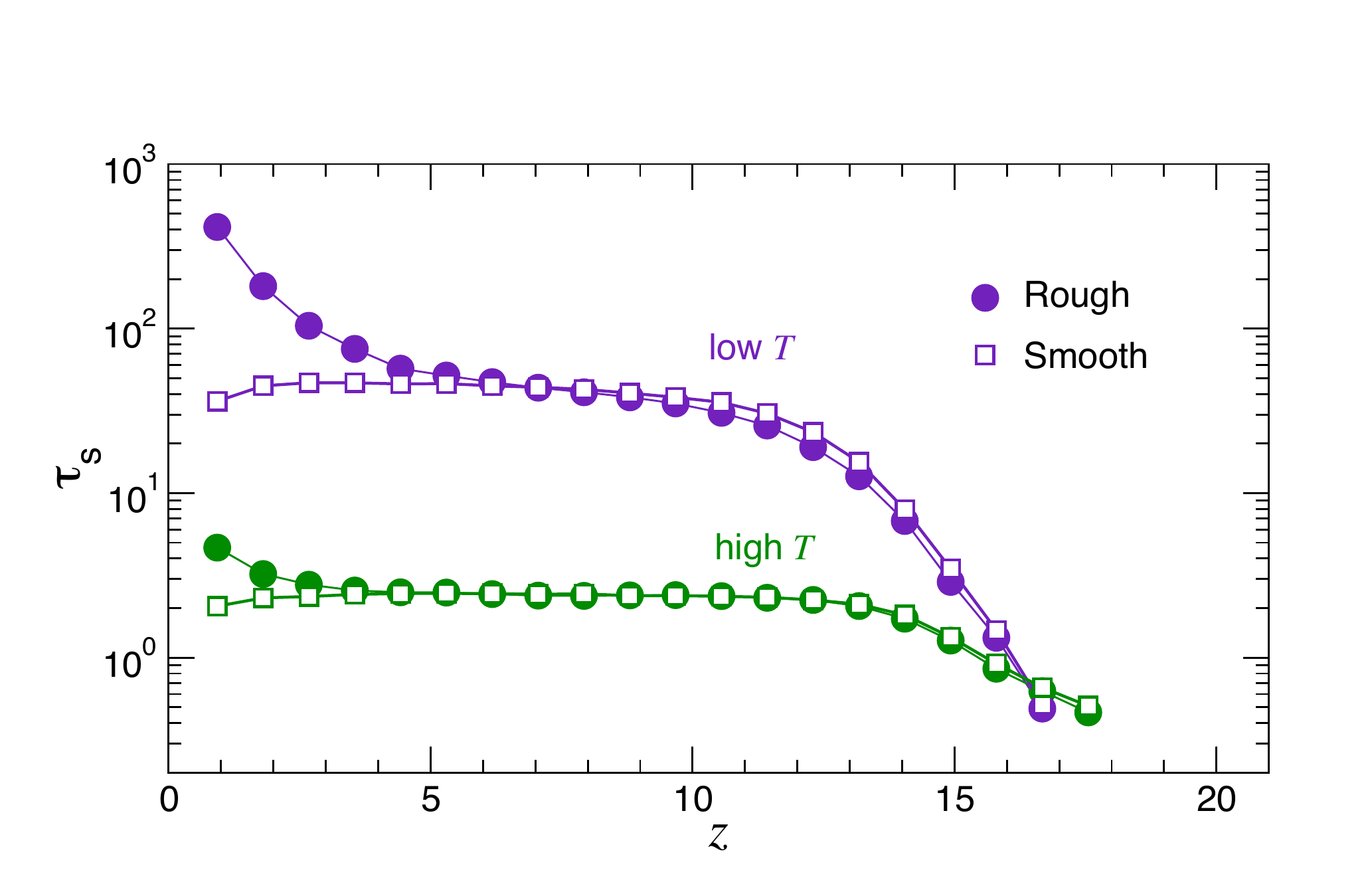}
\caption{ 
Relaxation time  $\tau_{\rm{s}}$ as function of distance  $z$ from the
substrate  Although   the  averaged  densities  of   two  systems  are
identical, the  local dynamics is  clearly distinct from  one another,
particularly  near the  the  substrate.}
\label{fig:local-all-rs-2}
\end{figure}

We next examine to what degree the local film
dynamics reflect the changes in the density described above. Figure~\ref{fig:local-all-rs-2} shows that the dynamics of
the film on a rough or smooth substrate at the same $T$ are nearly identical over the range from  the
center of the film to the free boundary region.  However, there are large
differences of relaxation time near the substrate.  The local
relaxation time $\tau_{\rm s}$ increases close to the rough substrate,
but decreases near the smooth substrate.  The enhanced dynamics near the
smooth substrate are in part a consequence of the fact the monomers can ``slide''
along the substrate due to the substrate smoothness (see Refs.~\cite{hanakata12,
  basch03}). This effect disappears for a rough substrate.  An
increasing relaxation time  approaching the rough substrate has also been observed in a computational study of
a binary Lennard-Jones liquids, as well as in a bead-spring model of
polymer melts with a relatively strong interaction,~\cite{scheidler02, basch05, smith03, virgiliis12}.  Evidently,
 substrate roughness is highly relevant for the polymer
film dynamics, and this factor must be controlled  for
consistent results.

\begin{figure}[ht]
\centering\includegraphics[trim=0mm 10mm 0mm 0mm, clip,width=0.48\textwidth]{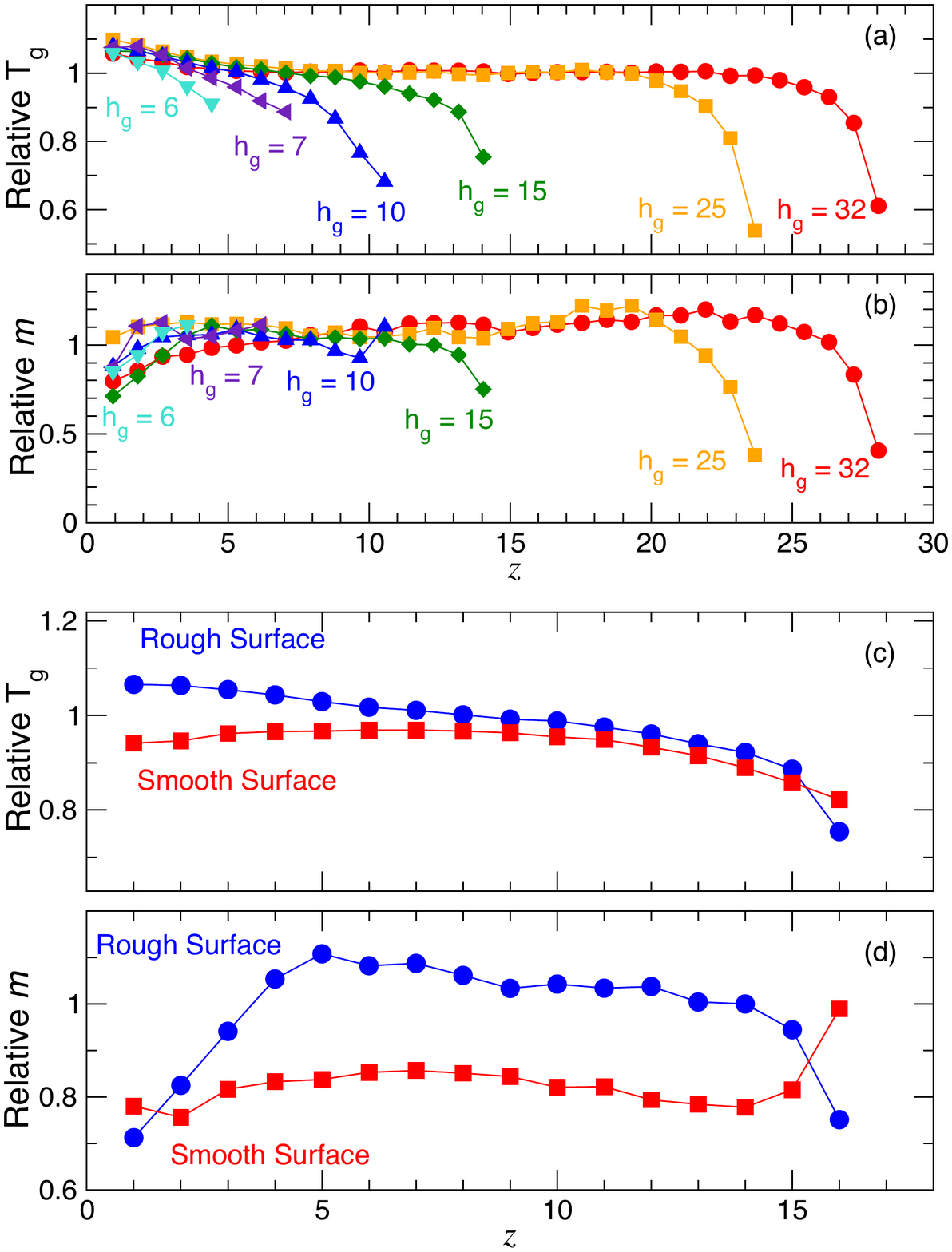}
\caption{{Local  $T_g$ and  fragility. }  (a) Relative  $T_g$ and  (b) $m$  of
polymer films supported  on rough substrate as function  of distance $z$
from the  substrate for  many thicknesses. (c) Relative  $T_g$ and (d)  $m$ of
a polymer  film ($h_g=15$) supported  on a rough  or a smooth  substrate. Near  the free
substrate $T_g$  decreases for both  system.  Near the  substrate, $T_g$
increases  for film  supported  on rough  substrate,  but decreases  for
smooth substrate. }
\label{fig:tg-m-layer}
\end{figure}

A  convenient  way  to  parameterize  local  dynamical  changes  is  by considering the local
dependence of $T_g$  and $m$ as function  of distance $z$ from
the  substrate.   This provides a way of summarizing  the behavior of $\tau_{\rm  s}(z,T)$,
shown in  Fig.~\ref{fig:local-all-rs-2}.   Figure~\ref{fig:tg-m-layer} (a)
shows that  $T_g$ increases near the  rough substrate,  reflecting the observed  increase of  $\tau_{\rm s}$  near the
attractive substrate.   Near the free boundary region, $T_g$  decreases due to
the enhanced mobility  of monomers at the free boundary region.  For
relatively  thick films, we find that there  is a  substantial film region where monomers have a $T_g$
close to the bulk value.  This is a situation in which the film thickness is large
compared      to      the      perturbing      scales      of      the
interfaces~\cite{hanakata12}. $T_g$ is  often found to be proportional
to $m$, as observed in the overall dynamics. However, we do not see
this proportionality  between the local $T_g$  and $m$.  Specifically,
$m$  {\em  decreases}  approaching   the  rough  substrate  while  $T_g$
increases.    This  opposing   trend   has  also   been  observed   in
polymer-nanoparticle composites~\cite{beatriz13}.

Figures~\ref{fig:tg-m-layer} (c) and (d) contrast the local variation
of $T_g$ and $m$ for rough and smooth substrates of a relatively thick
film, $h_g = 15$. In contrast to the increasing $T_g$ of polymer films
near the rough substrate, $T_g$ of smooth substrate decreases close to the smooth substrate, which is consistent with variation of
$\tau_{\rm s}$ (Fig.~\ref{fig:local-all-rs-2}).  Note that $T_g$ and
$m$ are slightly depressed for films on the smooth substrate, even at
the middle of the film, a scenario where the perturbing scales of both
interfaces become comparable to film thickness.

\section{Dependence of Dynamics on Substrate Strength and  Rigidity}
\begin{figure}[ht]
\centering\includegraphics[trim=0mm 0mm 0mm 5mm, clip,width=0.48\textwidth]{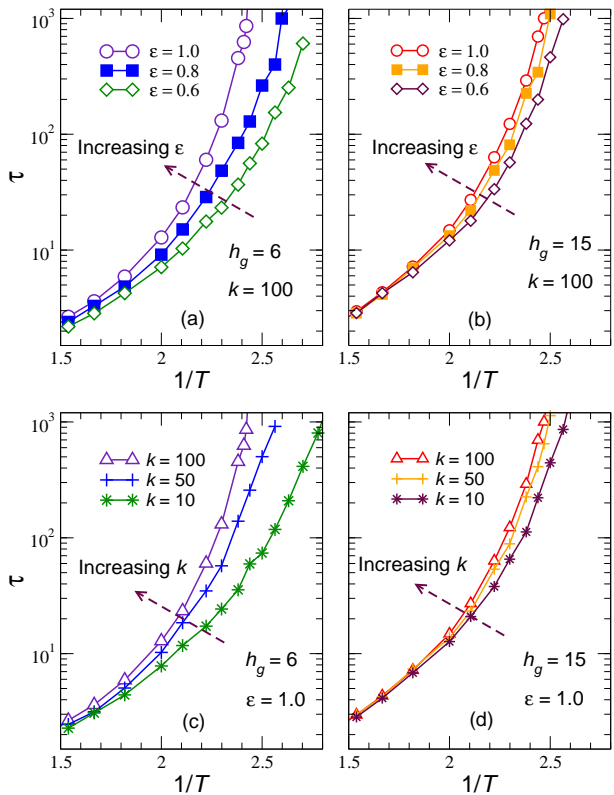}
\caption{The $T$ dependence of the relaxation time $\tau$ for two film
thicknesses,  $h_g$=15  and   6,  with  various  interfacial  strength 
$\varepsilon$ (a) and (b), and fixed rigidity $k=100$.  Panels (c) and (d) show the effect of variable rigidity k for fixed $\varepsilon=1$.  In general, $\tau$ is decreased as  we decrease the
substrate strength or the molecular-substrate stiffness.  }
\label{fig:relax_e-k}
\end{figure}
\subsection{Survey of Substrate Interaction and Rigidity Effects}

Substrate roughness is  relevant to the film dynamics, but
there are other crucial variables. We next
investigate the dependence of dynamics on the interaction
strength as well as rigidity of the rough substrate.  First, we
examine the role of substrate interaction strength.
Figures~\ref{fig:relax_e-k} (a) and ~\ref{fig:relax_e-k}  (b) show how relaxation time $\tau$ for
two representative film thicknesses changes as we vary the interaction
$\varepsilon$ between the rough substrate and the polymers.  The
overall changes in dynamics result from the competing effects of the
substrate and free interface, so that $\tau$ can be
higher or lower relative to the bulk.  As we have established, the
free boundary region decreases $\tau$ while a substrate with a
relatively strong interaction increases $\tau$.  Thus, for a
given thickness, $\tau$ decreases with decreasing the substrate
interaction strength.

We find  a similar effect  by varying the  stiffness $k$ of  the bonds
describing  the  substrate  stiffness.   Specifically,  increasing  the
flexibility  of the  substrate  (decreasing  $k$)  results in  a
smaller $\tau$  (Fig.~\ref{fig:relax_e-k} (c) and  (d)). Evidently,
monomers of the chains near the substrate are less constrained, since the substrate atoms
are more flexible.   The complete local analysis of the dependence of dynamics on flexibility of the substrate will
be    discussed     in    the    next    subsection.     By    comparing
Figs.~\ref{fig:relax_e-k} (a) and (b), as well as (c) and (d), we can
 see  that the  substrate interaction or  the flexibility  of the
substrate have greater influences  on the thinner film, expected since
the thinner film has a larger surface-to-volume ratio.
\begin{figure}[ht]
\centering\includegraphics[trim=0mm 5mm 0mm 5mm, clip,width=0.48\textwidth]{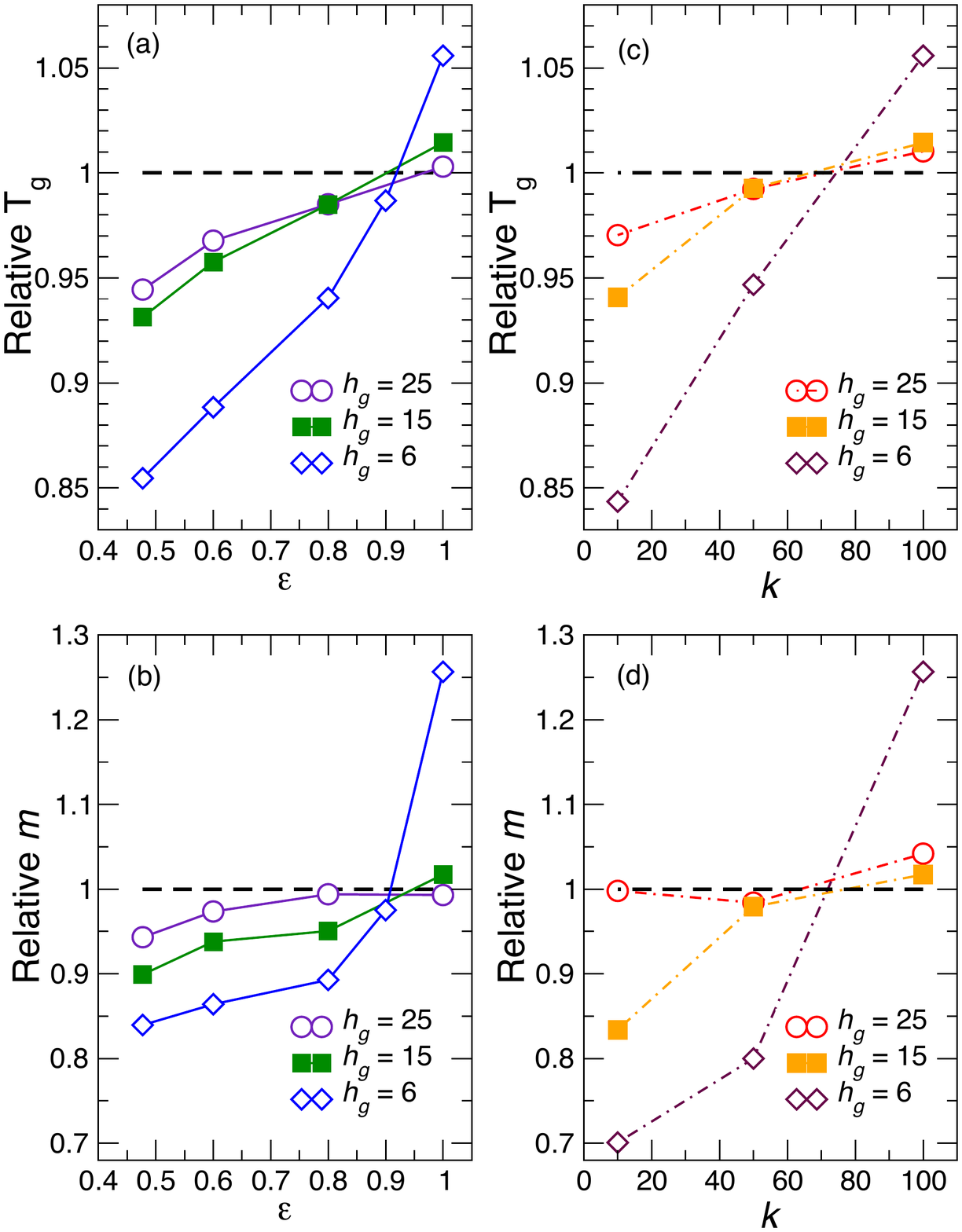}
\caption{Dependence   of  the   relative  $T_g$   and  $m$   of  three
  representative    film   thicknesses    as a function of     substrate   strength
  $\varepsilon$ (a) and (b)  or substrate rigidity $k$ (c) and
  (d). For thinner  films, the range of $T_g$ and $m$  is wide due to
  the larger substrate-to-volume ratio. }
\label{fig:tg-m_e-k}
\end{figure}

We next evaluate the resulting dependence of $T_g$ and fragility on
the substrate interaction strength and rigidity of the rough
substrate films.  Figures~\ref{fig:tg-m_e-k} (a) and (b) show how $T_g$ of
three representative thicknesses change as a function of substrate
interaction strength $\varepsilon$ at fixed rigidity $k = 100$.
Generally, increasing the polymer-substrate interaction increases both
$T_g$ and fragility as monomer dynamics near the substrate presumably
become progressively slower.  These general trends of a decreasing
$T_g$ with decreasing substrate interaction have also been observed
both in experiments and computational works~\cite{depablo00-2,
  depablo01, michels11, michels12}.  This depression of fragility is
also consistent with the findings in a free-standing
film~\cite{rigg-prl06}, which formally corresponds to taking the limit $\varepsilon\rightarrow 0$.  Evidently, the dependence of substrate polymer
interaction of $T_g$ or $m$ becomes more significant for thinner
films, as indicated by a steeper variation of $T_g$ or $m$ with
$\varepsilon$.

We found similar trends for $T_g$ and $m$ by varying the substrate
rigidity.  That is, increasing substrate rigidity $k$ at fixed substrate
interaction strength ($\varepsilon=1$) increases both $T_g$ and
$m$. It is interesting to note that there appears to be a nearly fixed
point for $T_g$ and $m$ as function of $\varepsilon$.
Specifically, $T_g$ and $m$ are independent of film
thickness for $\varepsilon\simeq0.9$ ($k=100$) or $k\simeq75$
($\varepsilon=1$). We emphasize that this does not mean there are no changes
in local dynamics, but rather that there is a balance between
the dynamic enhancement at the free boundary region and the slowing down of
the dynamics near the substrate. In fact, the increasing behavior of $m$  with decreasing thickness is only observed for values $k>75$ and $\varepsilon>0.9$. This 
 compensation effect is reminiscent of the self-excluded volume interactions of
polymers in solution near their $\theta$ point~\cite{Yamakawabook,freed85}, and the compensation point for isolated polymers interacting with surfaces~\cite{douglas89}.

Both results potentially  offer us insights into how  $T_g$ changes in
multilayer films, which are ``stacks'' of polymer films with different
species   characterized   by   different  flexibility,   inter-polymer
interaction, or molecular  weight.  Multilayer film
experiments by Torkelson and co-workers  have shown that a given layer
of the multilayer film may have different $T_g$ depending on the properties
of neighboring  layers~\cite{tork07}.  Here, we  emphasize that changes
in dynamics do not  necessarily arise from the substrate interaction
strength alone; changes in the  rigidity of the interface (e.g. polymer
films placed on a polymer substrate with the same substrate interaction strength, but having
different  molecular  flexibility) and substrate roughness  are also relevant.
\subsection{Local Structure and Dynamics}
\begin{figure}[ht]
\centering\includegraphics[width=0.48\textwidth]{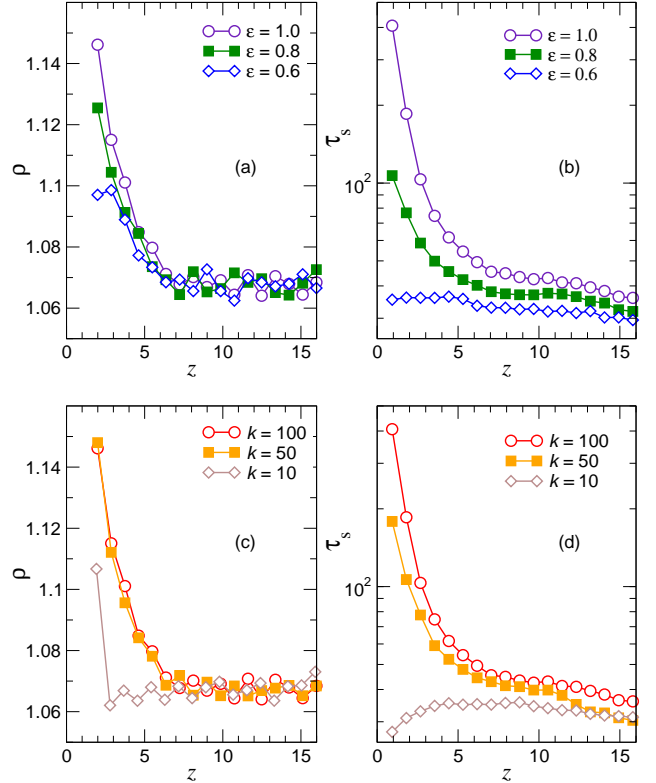}
\caption{Variation  monomer  of the density  profile  $\rho(z)$   and  local
relaxation  time $\tau_{\rm  s}$  as function  of distance  $z$ from  the
substrate  with varying  polymer-substrate  interaction $\varepsilon$ (a) and (b) 
or substrate rigidity $k$ (c) and (d). Note that $\rho$ becomes nearly constant for  $z>6$, but  for $z<6$, $\rho$  depends sensitively on the  interaction at the boundary.  }
\label{fig:dynamic-static_e-k}
\end{figure}
We revisit  our analysis of both film  structure and dynamics
(as in the previous subsection) to further confirm our arguments about the
role    of   substrate    changes   on    the    overall   dynamics.
Figures~\ref{fig:dynamic-static_e-k} (a) and (c) show how the monomer
density  $\rho(z)$  changes by  varying  the  substrate strength  or
rigidity  of the  substrate.  Far  from either  substrate, $\rho(z)$  has a
nearly constant value that is close to the bulk value.  In general,  the density near the
substrate  increases slightly,  as we  increase $\varepsilon$ or
$k$. The behavior of the  density at  very low rigidity ($k=10$) differs from more rigid substrates.
These  most flexible substrates
can be thought of as an amorphous solid that does not perturb the
film  much, because  it can  not  adapt its structure  to that  of  the polymer film.

Similar to our findings  comparing rough and smooth substrates,  Figures~\ref{fig:dynamic-static_e-k}(b) and (d) show
substantial changes in local relaxation $\tau_{\rm s}$ at the substrate
as function of $\varepsilon$ or $k$.
  Local relaxation time $\tau_{\rm
  s}(z)$ generally decreases as we decrease $\varepsilon$ or
$k$.  The weaker substrate interaction allows monomers to avoid caging
near the attractive substrate.  Likewise, decreasing substrate rigidity
allows monomers to move freely, since the substrate atoms are not strongly
localized. These dynamical changes do not mirror the
changes in the local density.
 This again emphasizes the limitations of a free volume based
interpretation of results..

\section{Decoupling and the `Fractional' Stokes-Einstein Relation}

One of the canonical features of glass-forming liquids is that the decoupling of viscous and diffusive relaxation processes gives rise to
a breakdown of the  Stokes-Einstein (SE) relation  approaching the glass transition.
This `decoupling' phenomenon is frequently associated with the emergence
of heterogeneity of the dynamics, which we know is prevalent in our
thin polymer films.  Normally, decoupling is quantified by the relation between
the diffusion coefficient $D$ and viscosity or a collective relaxation
time.  For polymer chains, $D$ is not readily accessible computationally since the mean-square displacements $\langle r^2(t) \rangle$ of the chain center of mass only reaches the diffusive
regime 
after extremely long times when the polymer melt is cooled.  Instead, reference~\cite{sds13} has
offered evidence that the characteristic time $t^*$ at which the non-Gaussian
parameter, 
\begin{equation}
\alpha_2(t)=\frac{3 \langle r^4(t) \rangle }{5 \langle r^2(t)
\rangle^2}-1, 
\end{equation}
has a maximum provides a diffusive relaxation time that exhibits a decoupling relation to the $\alpha$  relaxation time $\tau$ of the segmental dynamics.

\begin{figure}[ht]
\centering\includegraphics[trim=0mm 0mm 0mm 0mm, clip,width=0.48\textwidth]{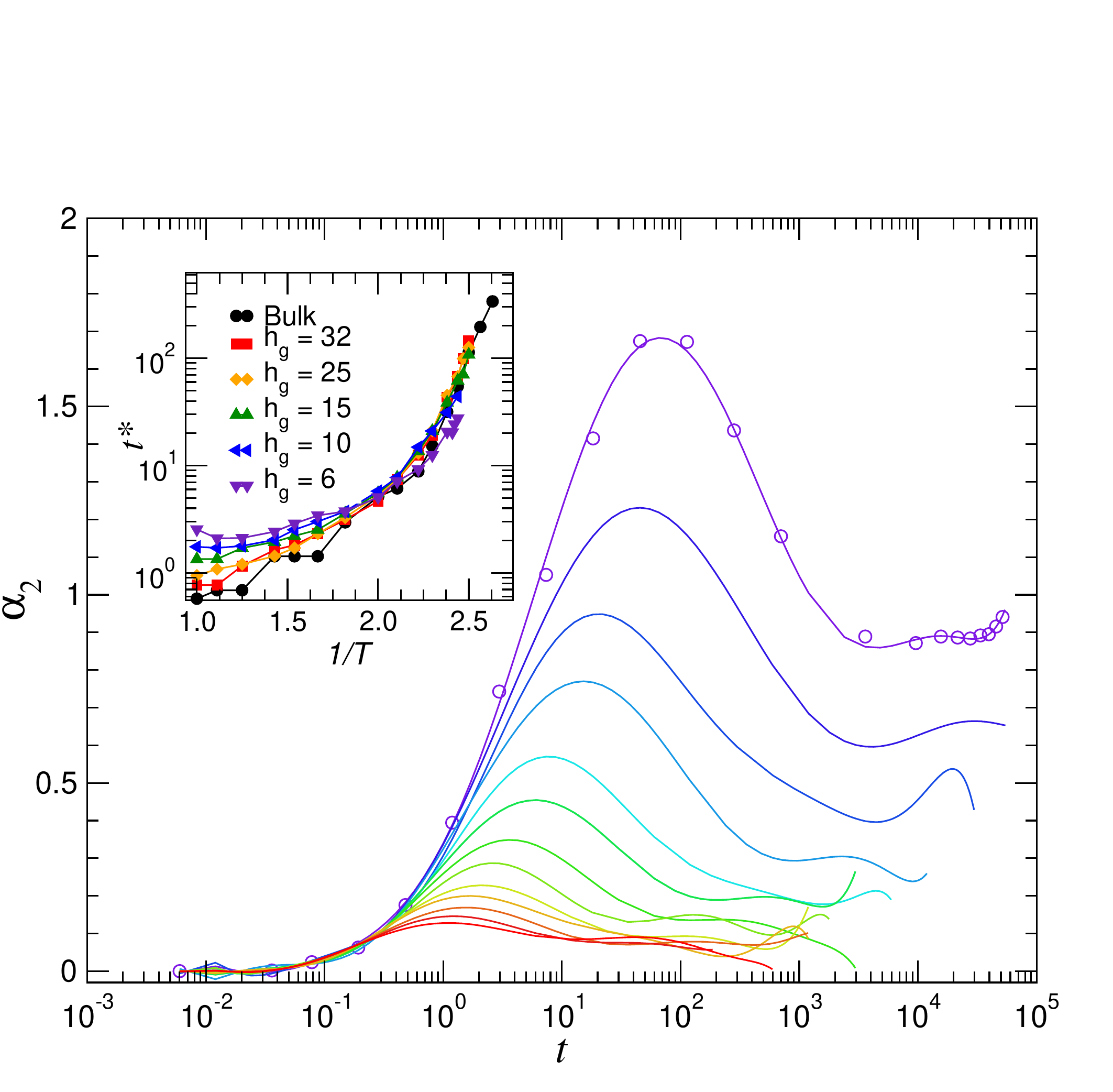}
\caption{The non-Gaussian parameter $\alpha_2(t)$ for various $T$ for
  the case of film thickness 15 with a rough, attractive substrate ($\varepsilon=1$, $k=100$).  The
  peak of $\alpha_2(t)$ defines the characteristic time scale $t^*$. The
  inset shows the $T$ dependence of $t^*$ for various film thicknesses.
  Data for other substrate interactions are qualitatively similar.}
\label{fig:alpha2}
\end{figure}

To broadly characterize the heterogeneity of segmental motion and
estimate a diffusive time scale, Fig.~\ref{fig:alpha2} shows
$\alpha_2(t)$ for many $T$ for $h_g=15$ with a rough or attractive
substrate.  Data for other thickness and substrate interactions show the
same qualitative features, so we only show this representative
example.  As is widely appreciated, $\alpha_2$ exhibits a peak at
intermediate time $t^*$.  One unusual feature of these data is
that $\alpha_2(t)$ does not decay to zero for large $t$, 
reflecting the fact that displacement perpendicular to the substrate is intrinsically limited by film thickness. The inset of Fig.~\ref{fig:alpha2} shows the $T$
dependence of $t^*$ for many film thicknesses for the rough substrate,
and it appears that $t^*$ grows less rapidly on cooling for
increasingly thin films.  Again, data for different substrate
interactions show the same general trends.

\begin{figure}[ht]
\centering\includegraphics[trim=0mm 0mm 0mm 0mm, clip,width=0.48\textwidth]{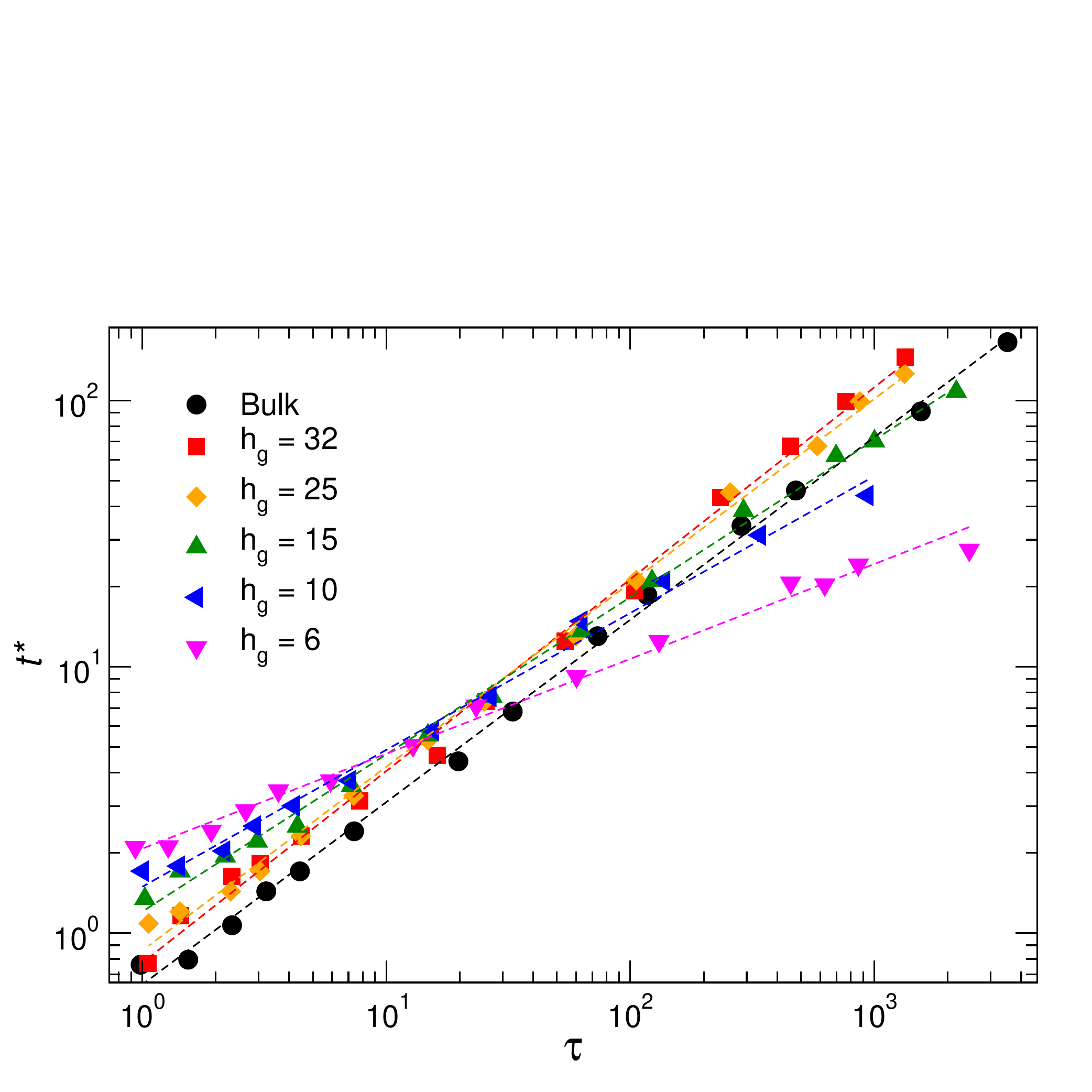}
\caption{Parametric relation between the coherent relaxation time $\tau$
  and $t^*$ for various film thickness for the representative case of a rough substrate with $k=100$ and $\varepsilon=1.0$,
  demonstrating a fractional power law relation $t^* \sim \tau^{1-\zeta}$.
  Clearly, $\zeta$ increases with decreasing film thickness as illustrated in Fig.~\ref{fig:decoup-fragility}.  }
\label{fig:fSE}
\end{figure}

We now use our data for $t^*$ and $\tau$ to quantify the decoupling of
relaxation time scales.  Typically, this decoupling gives rise to a
`fractional Stokes-Einstein' (fSE) relation described by a power scaling law,
\begin{equation}
t^* \sim \tau^{1-\zeta},
\label{eq:frac-se}
\end{equation}
where $\zeta<1$ is a fractional exponent characterizing the decoupling strength, so that $\zeta=0$ defines the simple case where the Stokes-Einstein relation is valid.  
Figure~\ref{fig:fSE} illustrates this variation for the representative case of polymer films on a rough  substrate with  $k=100$ and $\varepsilon=1.0$, where  $\zeta$ ranges from roughly 0.65  in the thinnest film to 0.3 upon approaching the bulk limit. 
A similar increase of the decoupling strength  $\zeta$ with increasing nanoparticle concentration has also been found in polymer nanocomposites for both attractive and non-attractive polymer-nanoparticle interactions~\cite{bea-new}.  

The decoupling phenomenon is an inherently more complicated problem in polymeric materials than for small molecule liquids because there are separate relaxation timescales for the segmental motions within the chains and for center of mass motions (which occur at a much longer time scales associated with the displacement of the chain as whole). 
Recent work has shown that the fragility of the segmental and overall chain motion relaxation processes are generally quite different~\cite{sokolov09Decoupling}. Moreover, Ediger and coworkers~\cite{Urakawa04} found a complete absence of decoupling between the center of mass diffusion and shear viscosity in unentangled polystyrene over a wide temperature range.  This situation is contrasted with relaxation at a segmental timescale where we observe  a power law relation between  $\tau$ and $t^*$.  
Sokolov and Schweitzer have separately considered a power law relation between  the segmental $\tau$ and polymer chain relaxation time, which they also described as being a `decoupling' relation~\cite{sokolov09Decoupling}. We do not attempt to describe this result because the calculation of the chain relaxation time at low temperatures is computationally prohibitive, and because we do not believe that the `decoupling' relationship of Sokolov and Schweitzer is analogous to the decoupling relation found in small molecule liquids. Of course, this relation between the segmental and large scale chain dynamics is fascinating and deserves further study.  

We next consider  experimental observations suggesting a direct relation between fragility and decoupling in small molecule glass formers.
Decoupling in glass-forming liquids has mainly been studied in context of crystal growth~\cite{Ediger06,ediger08Decoupling}, where the decoupling exponent $\zeta$ is inferred indirectly from the relationship  between the crystal growth cooling rate, and fluid viscosity. in particular Ref.~\cite{ediger08Decoupling}, shows a proportional relation between $m$ and $\zeta$  for low molecular weight organic and inorganic  glass-forming liquids for a wide variation in fragility~\cite{ediger08Decoupling}, although this work specifically excluded polymeric materials.  Sokolov and Schweizer~\cite{sokolov09Decoupling} studied the decoupling exponent relating the segmental relaxation time and collective chain motion relaxation time for different polymer glass forming materials based on dielectric relaxation time measurements, and found a monotonic increase $\zeta$  with fragility for this type of decoupling. Together, these works suggest that  the degree of decoupling  might generally increase in systems having larger fragility, although there is no generally accepted theoretical understanding of why such a relation might exist.

\begin{figure}
\centering\includegraphics[width=0.48\textwidth]{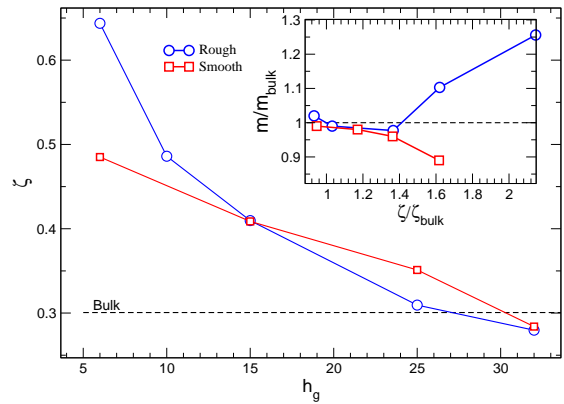}
\caption{Dependence of the decoupling exponent $\zeta$ on film thickness
  for rough and smooth substrate interactions when $k=100$ and $\varepsilon=1.0$.  For both cases, $\zeta$ increases on
  decreasing thickness regardless of the type substrate interaction.  The inset shows  the relationship between  relative $\zeta$ and  relative $m$ values obtained by altering the thickness of the film. }
\label{fig:decoup-fragility}
\end{figure}

We compared this trend to our results in Fig.~\ref{fig:decoup-fragility}, where we observe that the decoupling strength $\zeta$ increases with decreasing $h_g$ for smooth and rough substrates. The inset of  Fig.~\ref{fig:decoup-fragility} illustrates how $\zeta$ varies with $m$, and we see that the relation between $\zeta$ and $m$ does not follow a single trend in these two cases. A simple proportional relationship between $\zeta$ and $m$ does not describe our data, raising a question about the general relation between  decoupling strength and fragility.  
However, decoupling seems to be uniformly enhanced by geometrical confinement.
Sengupta et al.~\cite{sastry13} have observed 
 a diminished decoupling in small molecule liquids  between $D$ and  $\tau$ 
  with an increase of spatial dimensionality.
 If we view making polymer films thinner as reducing the `effective' spatial
dimension, then our observations fully accord with those of 
Sengupta et al. Simulations of polymer nanocomposites also show a
progressive reduction of fragility with nanoparticle concentration~\cite{beatriz13}, an effect
that might likewise be rationalized by an effective dimensional reduction with increased particle concentration, a point of view advocated previously~\cite{sausage}. While a definite relation between fragility and the strength of decoupling seems unlikely from the findings of 
Fig.~\ref{fig:decoup-fragility},  
dimensionality does seem to be relevant to understanding this effect.

\section{Collective Motion in Thin Polymer Films}

It has long been argued that polymer relaxation is governed by the scale of cooperative motion.  From a theoretical perspective, the classical
arguments of Adam-Gibbs (AG)~\cite{ag}, and our extension of this model based on numerical simulation evidence  and thermodynamic modeling~\cite{pds14}, provides a theoretical perspective  for testing this proposition. Specifically, according to AG theory, the activation Gibbs free
energy $\Delta G_a(T)$ is extensive of the size $z^*$ of `cooperatively
rearranging regions' (CRR), so that $\tau$ can be formally written in terms of the general transition state theory relation,
\begin{equation}
\tau(T)=\tau_0\,{\rm exp}[\Delta G_a(T)/\,k_{\rm B}T], \,   \Delta G_a(T)\equiv z^*\Delta \mu 
\label{eq:ag}
\end{equation}
where $\Delta \mu$ is the activation free energy at high temperatures when
particle motion does not involve a significant 
cooperative motion so that $z^*$ equals a constant (AG originally
assumed that $z^*\simeq1$ at high temperatures, corresponding to
completely uncooperative motion, but a constant value of $z^*$ at high $T$ is all
that is required to recover Arrhenius dynamics). Recent
simulations have shown that, despite the rather heuristic nature of the
original arguments of AG, Eq.~\ref{eq:ag} with $z^*$ identified
specifically with the average size $L$ of the cooperative string-like
particle exchange motion provides a good description for $\tau(T)$ in polymer melt simulations,
even in the case when nanoparticles have been added to tune the
fragility over a wide range~\cite{hanakata12, beatriz13,sds13,hds14,note1}. Very
recently, we have stablished a quantitative correspondence between the $L(T)$
and a living polymerization theory~\cite{pds14}, and inspired by these results,  
Freed~\cite{Freed14} has systematically derived Eq.~\ref{eq:ag} from transition state theory assuming that the transition states involve many-body transition events in the form of equilibrium polymers  with $z$ as the average string length. These results together provide a predictive theoretical framework for understanding the
dynamics of glass-forming liquids.

\begin{figure}[b!]
\centering\includegraphics[trim=0mm 0mm 0mm 0mm, clip,width=0.48\textwidth]{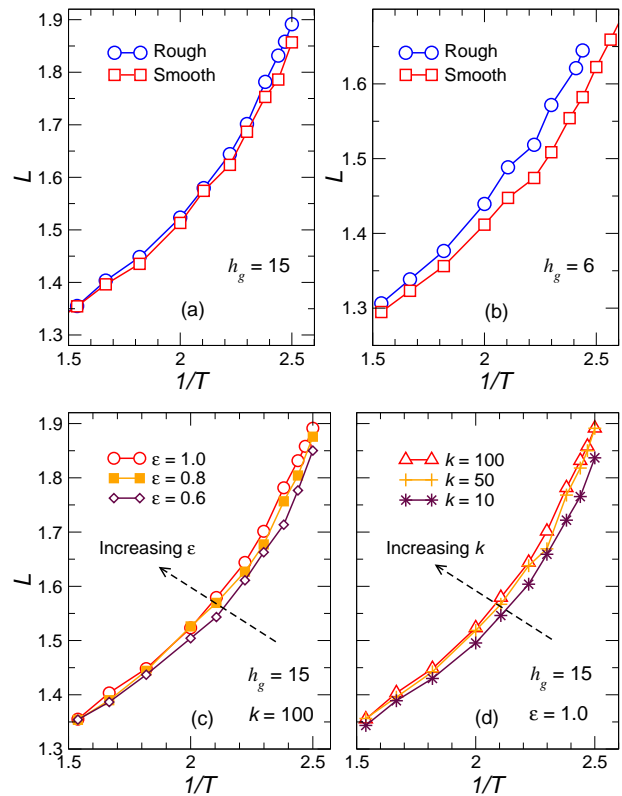}
\caption{Panels (a) and (b) show the comparison of $L(T)$ for films supported on a smooth or rough substrate. Panels (c) and (d) show the variation of t$L(T)$ with varying polymer-substrate interaction $\varepsilon$ or substrate rigidity $k$.  The variation of $L(T)$ with $h_g$, $\varepsilon$ and $k$ mimics that  of 
    $\tau(T)$ shown in Figures~\ref{fig:tau-tg-rs} and~\ref{fig:relax_e-k}. This
  qualitative consistency suggests the applicability of the string model of relaxation~\cite{pds14} to quantify $\tau(T)$.}
\label{fig:strings}
\end{figure}

We next evaluate $L(T)$ following methods
described in previous works~\cite{donati98, aichele03,sds13} to see if we can also describe the dynamics of thin polymer films within the same formalism. Figures~\ref{fig:strings}
(a) and (b) compare the $L(T)$ for two film thicknesses with 
a rough or smooth substrate at two representative
thicknesses. For a given film thickness, $L(T)$ is larger and grows faster
for a film with rough substrate, which is qualitatively
consistent with $\tau(T)$
(Fig.~\ref{fig:tau-tg-rs} (a) and (b)). The variation of
$L(T)$ with substrate interaction strength $\varepsilon$ or
substrate rigidity $k$, shown in Figs.~\ref{fig:strings} (c) and (d), and $L(T)$ in these cases also
 qualitatively captures the variation of $\tau(T)$ (Figs.~\ref{fig:relax_e-k} (b) and (d)). 
The similarities between the variation of $L(T)$ and $\tau(T)$ suggest
that we may be able to predict changes in fragility from the variation
of $L(T)$, as found in previous works~\cite{sd11, beatriz13, rigg-prl06,
  hds14}.  Accordingly, in the following section we consider this
possibility, and develop a framework to explain the variation of the
activation parameters relating $L$ and $\tau$.

\section{Collective Motions as an Organizing Principle for
Thin Films Dynamics}
\label{sec:col}

From the discussion of the preceding sections, it is clear that there are a
variety of factors that can alter the dynamics of thin polymer films,
including film thickness, roughness, the polymer-substrate 
interaction and the stiffness of the substrate.  Impurities introduced
from the film casting process and possible heterogeneity in the
substrate chemistry due to, e.g., substrate oxidation, are
also relevant~\cite{rigg-prl06}. These effects are all significant and
the observed changes of the film dynamics involves the convolution
of all these  variables.  We clearly need some organizing
principle to explain how all these factors influence the film
dynamics, and guide the development of polymer films
with rationally engineered properties.

In this section, we explore a perspective that allows us to obtain a unified
understanding of the diverse dynamical changes.  The Adam-Gibbs
perspective of the dynamics of glassy materials generally emphasizes the
importance of collective molecular motion  to understand
rates of structural relaxation, and the results of the previous section
indicate the promise of such an approach.  In order to quantitively test Eq.~\ref{eq:ag}, we follow Refs.~\cite{sd11, sds13,
  beatriz13, sdspb14}, and identify CRR size $z^*$ with the relative size $L/L_A$ of
string-like cooperative particle arrangements.  $L_A \equiv L(T_A)$ is the value of the string size at the temperature $T_A$, above which an Arrhenius law for $\tau(T)$ holds. To determine $T_A$, we use the same definition as in Ref.~\cite{pds14}.

The analysis of the dynamics of our thin polymer films is
then based on the string model for the dynamics of glass-forming liquids~\cite{pds14,pds14pnas}, in
which $\tau$ is described by the AG inspired relation,
\begin{equation}
 \tau(h,T)=\tau_0(h)\exp\bigg[\frac{L(T)}{L_A(h)}\frac{\Delta \mu(h,T)}{k_BT}\bigg],
\label{eq:ag2}
\end{equation}
where  $\Delta \mu(h,T)$ is the high temperature activation free energy for $T >T_A$, 
\begin{equation}
\Delta\mu(h,T)= \Delta H_a(h) -T\Delta S_a(h),
\label{eq:mu}
\end{equation}
where $\Delta H_a(h)$ and $\Delta S_a(h)$ are the enthalpic and the entropic contributions of the high $T$ activation free energy respectively. These basic energetic parameters vary with  film thickness $h_g$ and type of interaction ($\varepsilon$ and $k$). Note that Eq.~\ref{eq:ag2} for $T\geq T_A$ becomes, $\tau(h,T)=\tau_0\exp[(\Delta \mu(h,T)/k_BT]$, the typical activation form of transition state theory. In fact, Eq.~\ref{eq:ag2} at  $T_A$ implies that $\tau_0(h)$ is not a free parameter, but instead is determined by
\begin{equation}
\tau_0(h)=\tau_A(h)\exp[-\Delta \mu(h,T_A(h))/k_BT_A(h)],
\label{eq:tau0}
\end{equation} 
where $\tau_A\equiv\tau(T_A)$ so that $\Delta H_a$ and $\Delta S_a$ are the only undetermined parameters in Eq.~\ref{eq:ag2}. This relation was noted and tested in Ref.~\cite{pds14pnas}.
The string model prediction for the structural relaxation time of a film of thickness $h$ can then be formally written, 
\begin{equation}
\tau(h,T)=\tau_A(h)\exp\bigg[\frac{L(T)}{L_A(h)}\frac{\Delta \mu(h,T)}{k_BT}-\frac{\Delta\mu(h,T_A)}{k_BT_A}\bigg],
\end{equation}
 where,  $\Delta H_a(h)$ and $\Delta S_a(h)$ are the only   parameters  on which $\tau$ depends, just as in ordinary transition state theory for homogeneous fluids.
 
\begin{figure}[ht]
\centering\includegraphics[width=0.48\textwidth]{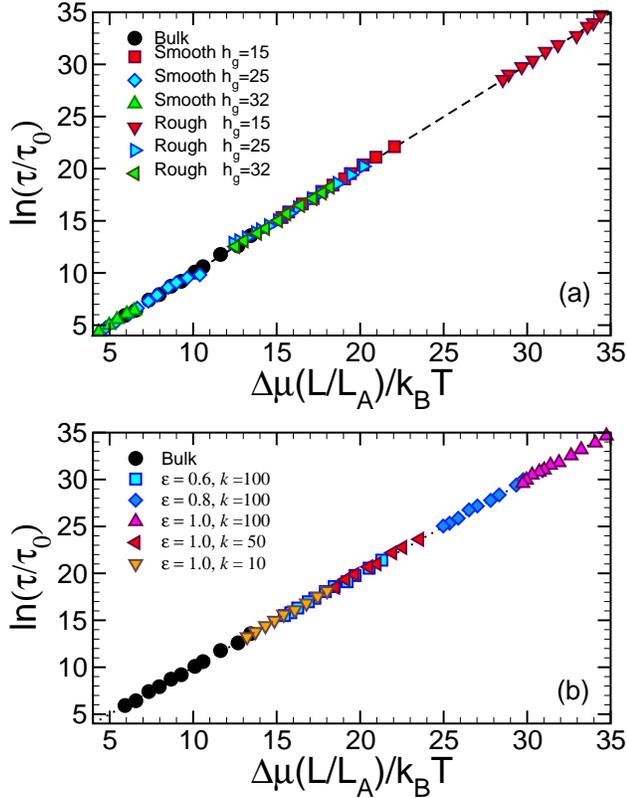}
\caption{Structural relaxation time $\tau$ in terms of the
average  strings size  $L$  for (a) various  thicknesses, and (b)  various
polymer-substrate interactions or substrate rigidities. $\tau$ is scaled by  
$\tau_0=\tau_A\exp[-\Delta \mu(T_A)/k_BT_A]$  
where  $\Delta \mu(T_A)=\Delta H_a -T_A\Delta S_a$, and $\Delta H_a$ and $\Delta S_a$ are determined by fitting to Eq.~\ref{eq:ag2} over a broad $T$ range.}
\label{fig:strings2}
\end{figure}

We now  demonstrate the applicability of
Eq.~\ref{eq:ag2} in quantitatively describing the dynamics of all the films we have thus far studied.
Figure~\ref{fig:strings2} shows the linear relationship between $\ln(\tau)$ and  $\Delta \mu L /k_BT$  for different films thickness in panel (a) and for different substrate rigidity and strength of interactions in panel (b). The universal collapse of $\tau$ in terms of string size was noted recently in a brief communications~\cite{hds14}, 
but the variation of the relaxation time prefactor  $\tau_0(h)$ was considered as a free parameter in that work.
We find that this data reduction holds for
all film thicknesses supported on a rough or smooth substrates, and applies as well as to the bulk polymer material. The same reduced variable description describes a representative
film supported on a rough substrate for various substrate interaction
or supporting substrate rigidity (see Fig.~\ref{fig:strings2} (b)). Although the data reduction is identical between these figures, we separate them
for clarity. This remarkable data reduction shows that we can quantitatively describe the film dynamics of all these films based on the string model relation (Eq.~\ref{eq:ag2}), despite a wide range
of dynamical changes due to film thickness, polymer-substrate
interaction, or substrate rigidity, 
For instance note that, Figs.~\ref{fig:strings}(c) and (d) show that the average extent of cooperative motion $L$ does not significantly change with $\varepsilon$ or $k$, but that  the  structural relaxation time does change considerably. Therefore, the changes in $\tau$ must result from  the variations of $\Delta H_a$ and $\Delta S_a$, which we next discuss.

\section{Theory for the Confinement effects on $\Delta S_a$ and $\Delta H_a$}

In order to develop a theoretical model of relaxation in polymer thin
films (or polymers with molecular additives and nanocomposites), we must understand what controls the basic activation parameters $\Delta
H_a$ and $\Delta S_a$ in the bulk polymer reference system.
In general, the activation parameters at high $T$ should depend on film
thickness, since all these properties depend on the film thermodynamic
properties. In this section, we specifically confront the issue of how the
parameters $\Delta H_a$ and $\Delta S_a$ depend on film thickness, boundary geometry, and interaction strength. We begin by
considering the variation of these energetic parameters in the high temperature
limit, where cooperative motion does not complicate our discussion.

\subsection{Transition State Theory}

Classical transition state theory~\cite{eyring1936,
  ewell1938, eyring1941-RateProc} implies that the diffusion
coefficient, the  structural relaxation time, and shear viscosity can all be
described by an Arrhenius $T$ dependence (at high $T$, where relaxation is not cooperative); i.e., the structural
relaxation time $\tau$ can be expressed by the Arrhenius expression
\begin{equation}
  \tau=\tau_0\exp[{\Delta G_a/k_{\rm
      B}T}],                
\label{eq:TST}
\end{equation}
where $\tau_0$ is the vibrational time, and the activation free energy $\Delta G_a$ is associated with the
displacement of a polymer statistical segment~\cite{eyringKuzman1940,
  eyring1958}.  As discussed earlier, the activation free energy, $\Delta G_a(T>T_A)\equiv\Delta \mu$, has  enthalpic contributions $\Delta H_a$ related to the strength of the intermolecular
cohesive interactions, and entropic contributions
$\Delta S_a$ arising from entropy changes needed to surmount complex multidimensional potential energy
barriers in condensed materials~\cite{bondi1946,
  yelon06, yelon11, yelon1990}.  Predicting $\Delta S_a$ is often a weak point in transition state modeling and the factor $\exp[-\Delta S_a/k_B]$ is often just absorbed into the measured prefactor $\tau_0$ as a practical matter, but this is not an option for glass-forming liquids.

To guide our thinking, we need to recognize the
physical origin for the values of $\Delta H_a$ and $\Delta S_a$. In order to understand  qualitatively  $\Delta H_a$, we go back to Eyring's early transition state theory
arguments~\cite{eyring1941-RateProc}, and consider long-standing
physical observations in simple fluids~\cite{madge1934, eirich1939,
  qunfang1997} that relate $\Delta H_a$ to the energy change
associated with the removal of a test molecule from its local environment in the fluid state~\cite{Madge34,Simha39,glasstone1941theory}. Such, an interpretation has recently been implemented computationally by Egami and coworkers~\cite{Egami13}. This perspective implies that
$\Delta H_a$ should scale in approximate proportion to the heat of vaporization $H_{\rm
  vap}$ or the cohesive interaction energy of the fluid. Although this argument is simple, its experimental validity  has been established for  hundreds of fluids~\cite{QunFang97}. More
recently, simulations of simple Lennard-Jones fluids in 2 and 3 dimensions have shown that
$\Delta H_a$ scales in proportion to the interaction parameter
$\varepsilon$~\cite{speedy1989,
  hentschel12, iwashita13}, the natural measure of intermolecular interaction strength in simple pair potential models such as LJ fluids and also our polymer model.  

As noted above, the variation of $\Delta S_a$ with molecular
parameters is less well understood. In many small molecule
fluids,  the intermolecular potential is weak, and therefore the variation of $\Delta S_a$  can be reasonably neglected. However, for molecules with
many internal degrees of freedom, such as polymers, there can be a
considerable variation in $\Delta S_a$. In particular, a survey by
Bondi~\cite{bondi1946} revealed that $\Delta S_a/k_B$ could vary over a
100 units, and can even change sign, so that  variations of
$\Delta S_a$ cannot be ignored. Bondi's pioneering study describes the
frustrations of early theoretical efforts to estimate $\Delta S_a$
theoretically for complex fluids.

\begin{figure}[ht]
\centering\includegraphics[width=0.48\textwidth]{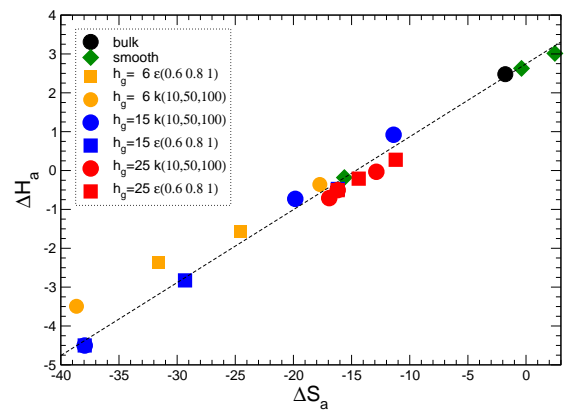}
\caption{Variation of $\Delta H_a$ and $\Delta S_a$ for
  different film thicknesses, substrate roughnesses, substrate
  interactions, and substrate rigidities. The slope defines compensation
  temperature $T_{\rm comp } = 0.18$. The data for the thinnest film ($h_g = 6$) shows a deviation from the thicker film data, which is discussed in the main text. }
\label{fig:HS}
\end{figure}

The
basic physical picture that the free energy of activation is related to
the free energy cost of removing a molecule from its local molecular
environment suggests a proportionate contribution to $\Delta S_a$ from the
cohesive intermolecular interaction~\cite{yelon1990, boisvert1998}. This
effect is evident in Trouton's rule~\cite{digilov04, nash1984}, which
relates the heat of vaporization $H_{\rm vap}$ and the entropy of
vaporization $S_{\rm vap}$ of gases and the Barclay-Butler
phenomenological relation linking enthalpies and entropies of solvation
in many mixtures~\cite{barclay1938, bell1937, evans1936, henry1945}.
Indeed, many studies have established the
specific relation ~\cite{yelon06, yelon11, yelon1990},
\begin{equation}
\Delta S_a = \Delta S_{\rm 0} + \Delta H_a / k_{\rm B} T_{\rm comp}                                                           
\label{eq:entropy-enthalpy}
\end{equation}
supported by observations on diverse
materials, where $\Delta S_{\rm 0}$ captures a background  contribution associated with the internal configurational degrees of freedom of fluid molecules. This linear relation has long been established for the
Arrhenius activation parameters of bulk polymer fluids~\cite{barrer1943,
  waring1947} . In glass-forming materials, the entropy-enthalpy
`compensation temperature' $T_{\rm comp}$ is often found to be near the
glass transition temperature of the fluid sometimes termed, the ``melting temperature of
the glass''~\cite{dyre1986}; in crystalline solid materials, $T_{\rm
  comp}$ is often found to be near the melting temperature $T_m$ of the
solid~\cite{eby62,dienes50}. We indeed find  `entropy-enthalpy' compensation in our simulated glass-forming films (shown in Fig.~\ref{fig:HS}) where the compensation temperature,  $T_{\rm comp}=0.18$, a temperature similar to the estimated VFT temperature. These observations suggest that $T_{\rm comp}$ is determined by a
 physical condition at which the intermolecular cohesive interaction
is insufficient to keep the material in the solid state, so that the
fluid then begins to explore liquid-like configurations, but a
quantitative understanding of how $T_{\rm comp}$ relates to the structure of the
potential energy substrate remains to be determined.
We next examine how $\Delta H_a$ and $\Delta S_a$ of transition state theory depend on the scale of confinement in thin polymer films. 
\subsection{Effect of confinement on $\Delta H_a$ and $\Delta S_a$}

\begin{figure}[htbp]
   \centering
   \includegraphics[width=0.48\textwidth]{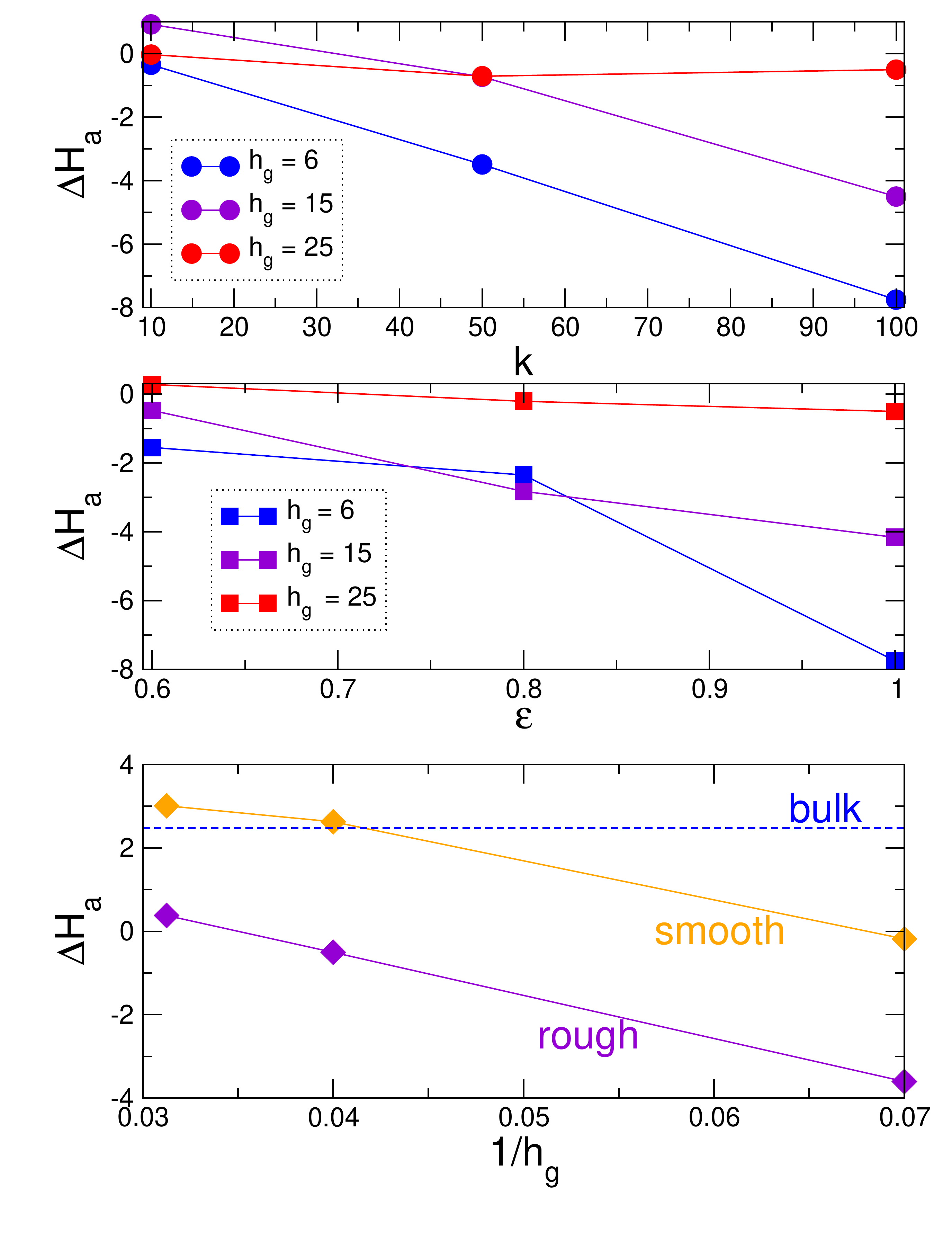} 
   \caption{Changes in the enthalpic contribution  $\Delta H_a$ of the activation free energy. (a) Effect of varying the roughness by varying  $k_{\rm s}$. (b) Effect of varying the strength of the substrate interaction by varying  $\varepsilon$. (c) Changing confinement on smooth or rough films.  }
   \label{fig:H}
\end{figure}

\begin{figure}[htbp]
   \centering
   \includegraphics[width=0.48\textwidth]{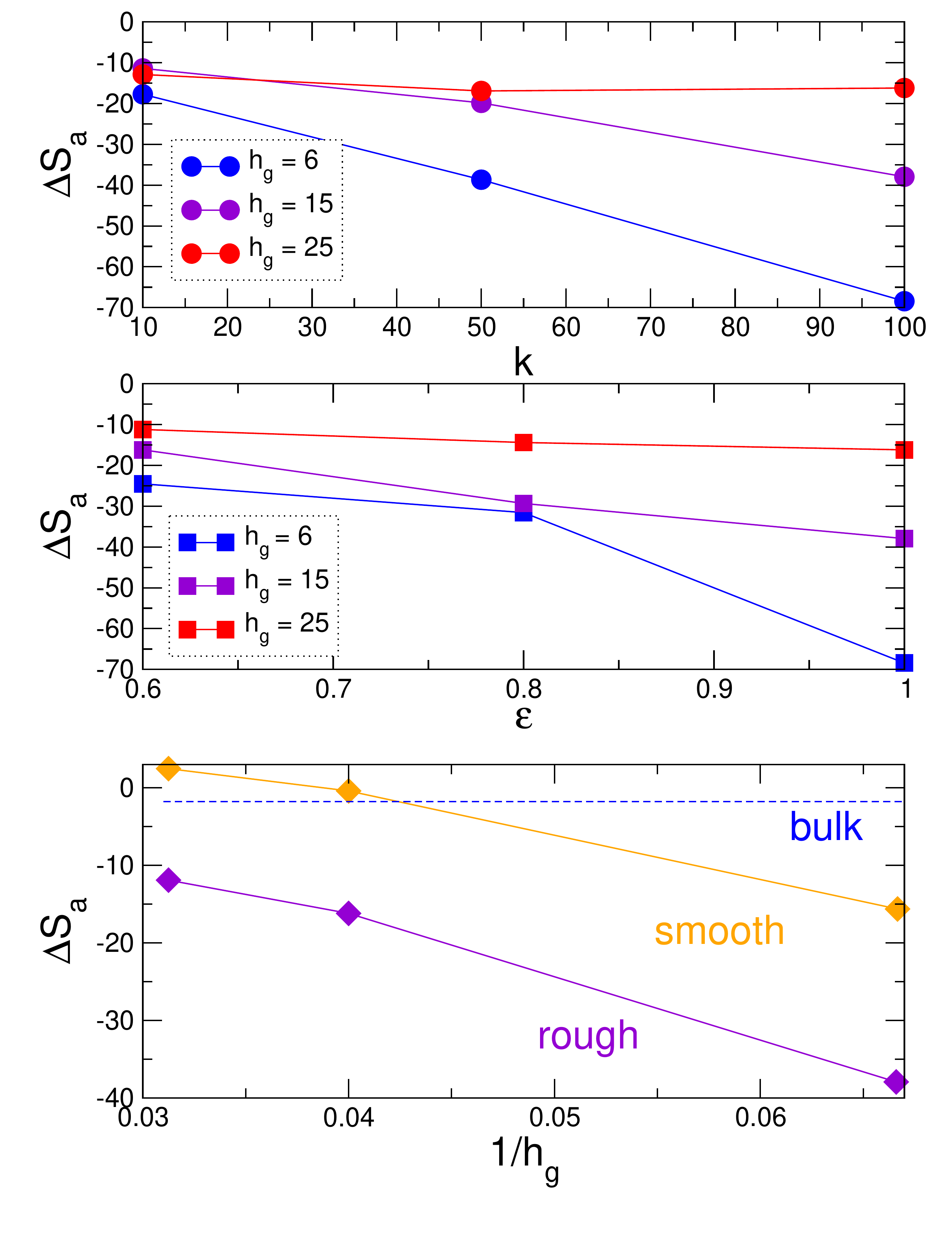} 
   \caption{Changes in the entropic  contribution $\Delta S_a$ of the activation  free energy. (a) Effect of varying the  roughness by varying the parameter $k_{\rm s}$. (b) Effect of varying the strength of the substrate interaction by varying the parameter $\varepsilon$. (c) Changing confinement on smooth or rough films.  }
   \label{fig:S}
\end{figure}

In general, film confinement (or the addition of additives to a fluid)
must change the effective molecular coordination number, and thus  the cohesive interaction strength, so that  $\Delta S_a(h)$ and
 $\Delta H_a(h)$ of activation  also vary with film thickness and boundary
interactions. 
This change in the effective coordination number due to changes in the surface-to-volume ratio with confinement has been extensively discussed in connection to estimate critical point shifts in confined fluids and magnetic materials~\cite{ff87,allan70},
and  as the origin of the Gibbs-Thomson effect for the shift of the melting point of crystals~\cite{mackena90}. This scaling argument suggests that $\Delta H_a(h)$
scales with the substrate-to-volume ratio of the film, so that 
\begin{equation}
 \Delta H_a(h)-\Delta H_a({\rm bulk)} \sim 1/h.
\label{eq:enthalphy-scaling}
\end{equation}
We  see that this scaling is roughly consistent with the data in Fig.~\ref{fig:H}, but varying the substrate energy of the film can also lead to a change in $\Delta H_a$, whose sign  depends on whether substrate-polymer interactions are
stronger or weaker than the polymer-fluid interactions.  In particular, Fig.~\ref{fig:H} shows that $\Delta H_a$ decreases with $k$ and $\varepsilon$.
Thus, there are a number of contributing substrate terms that influence $\Delta H_a$ so a clean $\sim1/h$  finite size scaling of $\Delta H_a$ is not obtained. Nonetheless, the general trend in variation of $\Delta H_a$ with confinement is understandable. We emphasize that despite the highly variable nature of the variation of $\Delta H_a$ and $\Delta S_a$ with boundary interaction, roughness, substrate rigidity and film thickness, the compensation relation between these activation energies holds to an excellent approximation. However, there is a different relation between $\Delta H_a$ and $\Delta S_a$ for film thickness $h_g= 6$, an effect that  is also reflected of the increase in the decoupling exponent $\zeta$ in Fig.~\ref{fig:fSE}.
This interesting effect requires further study, but it may be a consequence of the strong deviation of $\rho$ from its bulk  value very near the substrate. In particular, Fig.~\ref{fig:dynamic-static_e-k} shows large density gradients for $z<6$, so that  `ultrathin' films with $h_g \lesssim 6$ are rather unlike the bulk material and thicker polymer films. 
The component of $\Delta S_a$ that is linked to $\Delta H_a$ is expected from entropy-enthalpy compensation to follow the same dependence as $\Delta H_a$,
 suggesting a similar inverse dependence on thickness.  Fig.~\ref{fig:S}(c) shows this is the case. 

\begin{figure}
\centering\includegraphics[width=0.48\textwidth]{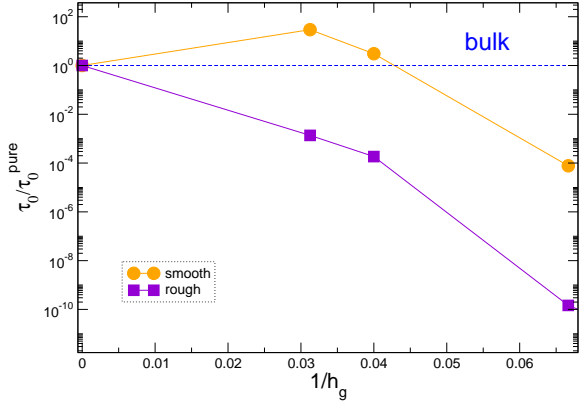}
\caption{Changes in  $\tau_0$ as a function of inverse of film thickness relative to the value for the bulk system. as a function of the inverse of the thickness $h_g$. The circles represent smooth substrate and the squares represent those for a rough substrate. }
\label{fig:tau0}
\end{figure}

We have not previously discussed the relaxation time prefactor $\tau_0$ 
in Eq.~\ref{eq:ag2}, which varies appreciably with confinement, as shown in Fig.~\ref{fig:tau0} for thin polymer films on smooth or rough substrates. In particular, Fig.~\ref{fig:tau0} indicates that the changes in $\tau_0$ can be as large as \emph{10 orders of magnitude}, so this factor is highly relevant for understanding  the changes in relaxation time in thin films and  nanocomposites. We again emphasize that $\tau_0$ is not a free parameter in the string model of glass-formation, but this quantity is entirely determined from $\Delta H_a$ and $\Delta S_a$ (Eq.~\ref{eq:tau0}).

 Although $\tau_0$ varies strongly with confinement, the relaxation time at $T_A$, $\tau_A$ varies only weakly, so that $\tau_0$ changes are due almost entirely to changes of $\Delta H_a(h)$ and $\Delta S_a(h)$.  This phenomenon has not been appreciated before.
 Evidently, the significant changes in the relaxation of glass-forming films and nanocomposites derive in large part from the high temperature activation parameters, which are typically thought not to have a direct relation to glass formation.

\section{Conclusions}
We have systematically explored  factors to
that alter the dynamics of thin supported polymer films-- film
thickness, substrate roughness, polymer-substrate interaction strength,
and the rigidity of the supporting substrate. All these factors were
found to be highly relevant to the dynamics of our simulated polymer
films and their coupling makes an understanding of changes in polymer
film dynamics in thin films rather complicated. Simple free volume
ideas are  inadequate to explain the significant
changes that we observe.  Control of boundary roughness and polymer-substrate interaction is  evidently necessary to make polymer
films with reproducible properties and the prediction of film
properties based on computation will require the specification of many
factors related to the film boundary conditions and structure.

Despite the wide variation of film dynamics with boundary
conditions and film thickness, we find that we can obtain a
remarkably general characterization of the changes in the film
dynamics for all  film conditions using the string model
of structural relaxation.  In particular, we quantitatively describe the change in dynamics of supported polymer films based
on how the collective motion is perturbed in the film. These changes are ultimately parameterized in terms of the high temperature activation free energy, leading to the almost paradoxical finding that glass formation is controlled by the fluid properties in the high temperature limit.

Given the computational success of the string model, we are now faced with the problem of determining the extent of string-like
collective motion in real materials.  Future work must address how the
extent of collective motion can be effectively estimated from direct
measurement.
Our recent measurements~\cite{pds14} suggest a direct relation between the string length and the interfacial mobility scale near the polymer-air-boundary of supported films and offers a promising method for estimating the string length. Moreover, recent work~\cite{zhang13} also suggests that noise measurements
might be effective for estimating average string length, and we
plan to pursue this possibility in the future.

 As a secondary consideration, we examined how the large changes in fragility in our simulations were related to changes between a diffusive relaxation time and the decoupling relation in the segmental relaxation time. We find that decoupling and fragility can change in opposing directions, calling into question a general relation between decoupling and fragility.  However, the changes in the decoupling exponent $\zeta$ are in accord with simulation results~\cite{sastry13} and theoretical arguments~\cite{sausage} for the effect of reduced dimensionality on decoupling. Thus, our findings do not support previous observations indicating a proportional change between fragility and the strength of decoupling in small molecule liquids, and we conclude that the relation between fragility and decoupling requires further study. Our observations also imply that it would be worthwhile to define the concept of effective dimensionality more precisely and check this perspective with changes of the decoupling strength. 

\bibliography{paper1}

\begin{thebibliography}{97}
\expandafter\ifx\csname natexlab\endcsname\relax\def\natexlab#1{#1}\fi
\expandafter\ifx\csname bibnamefont\endcsname\relax
  \def\bibnamefont#1{#1}\fi
\expandafter\ifx\csname bibfnamefont\endcsname\relax
  \def\bibfnamefont#1{#1}\fi
\expandafter\ifx\csname citenamefont\endcsname\relax
  \def\citenamefont#1{#1}\fi
\expandafter\ifx\csname url\endcsname\relax
  \def\url#1{\texttt{#1}}\fi
\expandafter\ifx\csname urlprefix\endcsname\relax\def\urlprefix{URL }\fi
\providecommand{\bibinfo}[2]{#2}
\providecommand{\eprint}[2][]{\url{#2}}

\bibitem[{\citenamefont{Bratton et~al.}(2006)\citenamefont{Bratton, Yang, Dai,
  and Ober}}]{litho-rev06}
\bibinfo{author}{\bibfnamefont{D.}~\bibnamefont{Bratton}},
  \bibinfo{author}{\bibfnamefont{D.}~\bibnamefont{Yang}},
  \bibinfo{author}{\bibfnamefont{J.~Y.} \bibnamefont{Dai}}, \bibnamefont{and}
  \bibinfo{author}{\bibfnamefont{C.~K.} \bibnamefont{Ober}},
  \bibinfo{journal}{Polym. Adv. Tech.} \textbf{\bibinfo{volume}{17}},
  \bibinfo{pages}{94} (\bibinfo{year}{2006}).

\bibitem[{\citenamefont{Hutmacher}(2001)}]{tissue-eng}
\bibinfo{author}{\bibfnamefont{D.~W.} \bibnamefont{Hutmacher}},
  \bibinfo{journal}{J. Biomat. Sci. Polym. Ed.} \textbf{\bibinfo{volume}{12}},
  \bibinfo{pages}{107} (\bibinfo{year}{2001}), ISSN \bibinfo{issn}{0920-5063}.

\bibitem[{\citenamefont{Stafford et~al.}(2006)\citenamefont{Stafford, Vogt,
  Harrison, Julthongpiput, and Huang}}]{stafford06}
\bibinfo{author}{\bibfnamefont{C.~M.} \bibnamefont{Stafford}},
  \bibinfo{author}{\bibfnamefont{B.~D.} \bibnamefont{Vogt}},
  \bibinfo{author}{\bibfnamefont{C.}~\bibnamefont{Harrison}},
  \bibinfo{author}{\bibfnamefont{D.}~\bibnamefont{Julthongpiput}},
  \bibnamefont{and} \bibinfo{author}{\bibfnamefont{R.}~\bibnamefont{Huang}},
  \bibinfo{journal}{Macromolecules} \textbf{\bibinfo{volume}{39}},
  \bibinfo{pages}{5095} (\bibinfo{year}{2006}),
  \urlprefix\url{http://dx.doi.org/10.1021/ma060790i}.

\bibitem[{\citenamefont{O'Connell et~al.}(2008)\citenamefont{O'Connell,
  Hutcheson, and McKenna}}]{connell08}
\bibinfo{author}{\bibfnamefont{P.~A.} \bibnamefont{O'Connell}},
  \bibinfo{author}{\bibfnamefont{S.~A.} \bibnamefont{Hutcheson}},
  \bibnamefont{and} \bibinfo{author}{\bibfnamefont{G.~B.}
  \bibnamefont{McKenna}}, \bibinfo{journal}{J. Polym. Sci. B}
  \textbf{\bibinfo{volume}{46}}, \bibinfo{pages}{1952} (\bibinfo{year}{2008}),
  \urlprefix\url{http://dx.doi.org/10.1002/polb.21531}.

\bibitem[{\citenamefont{Alcoutlabi and McKenna}(2005)}]{mckenna-rev05}
\bibinfo{author}{\bibfnamefont{M.}~\bibnamefont{Alcoutlabi}} \bibnamefont{and}
  \bibinfo{author}{\bibfnamefont{G.~B.} \bibnamefont{McKenna}},
  \bibinfo{journal}{J. Phys Condens. Matter} \textbf{\bibinfo{volume}{17}},
  \bibinfo{pages}{R461} (\bibinfo{year}{2005}), ISSN \bibinfo{issn}{0953-8984}.

\bibitem[{\citenamefont{Binder et~al.}(2008)\citenamefont{Binder, Horbach,
  Vink, and De~Virgiliis}}]{confinement-rev}
\bibinfo{author}{\bibfnamefont{K.}~\bibnamefont{Binder}},
  \bibinfo{author}{\bibfnamefont{J.}~\bibnamefont{Horbach}},
  \bibinfo{author}{\bibfnamefont{R.}~\bibnamefont{Vink}}, \bibnamefont{and}
  \bibinfo{author}{\bibfnamefont{A.}~\bibnamefont{De~Virgiliis}},
  \bibinfo{journal}{Soft Matter} \textbf{\bibinfo{volume}{4}},
  \bibinfo{pages}{1555} (\bibinfo{year}{2008}), ISSN \bibinfo{issn}{1744-683X}.

\bibitem[{\citenamefont{Keddie et~al.}(1994)\citenamefont{Keddie, Jones, and
  Cory}}]{keddie1}
\bibinfo{author}{\bibfnamefont{J.~L.} \bibnamefont{Keddie}},
  \bibinfo{author}{\bibfnamefont{R.~A.~L.} \bibnamefont{Jones}},
  \bibnamefont{and} \bibinfo{author}{\bibfnamefont{R.~A.} \bibnamefont{Cory}},
  \bibinfo{journal}{Europhys. Lett.} \textbf{\bibinfo{volume}{27}},
  \bibinfo{pages}{59} (\bibinfo{year}{1994}).

\bibitem[{\citenamefont{Forrest et~al.}(1997)\citenamefont{Forrest,
  Dalnoki-Veress, and Dutcher}}]{fdd97}
\bibinfo{author}{\bibfnamefont{J.~A.} \bibnamefont{Forrest}},
  \bibinfo{author}{\bibfnamefont{K.}~\bibnamefont{Dalnoki-Veress}},
  \bibnamefont{and} \bibinfo{author}{\bibfnamefont{J.~R.}
  \bibnamefont{Dutcher}}, \bibinfo{journal}{Phys. Rev. E}
  \textbf{\bibinfo{volume}{56}}, \bibinfo{pages}{5705} (\bibinfo{year}{1997}),
  ISSN \bibinfo{issn}{1063-651X}.

\bibitem[{\citenamefont{Fukao and Miyamoto}(2001)}]{fukao01}
\bibinfo{author}{\bibfnamefont{K.}~\bibnamefont{Fukao}} \bibnamefont{and}
  \bibinfo{author}{\bibfnamefont{Y.}~\bibnamefont{Miyamoto}},
  \bibinfo{journal}{Phys. Rev. E} \textbf{\bibinfo{volume}{64}},
  \bibinfo{pages}{011803} (\bibinfo{year}{2001}),
  \urlprefix\url{http://link.aps.org/doi/10.1103/PhysRevE.64.011803}.

\bibitem[{\citenamefont{Fryer et~al.}(2001)\citenamefont{Fryer, Peters, Kim,
  Tomaszewski, de~Pablo, Nealey, White, and Wu}}]{depablo01}
\bibinfo{author}{\bibfnamefont{D.~S.} \bibnamefont{Fryer}},
  \bibinfo{author}{\bibfnamefont{R.~D.} \bibnamefont{Peters}},
  \bibinfo{author}{\bibfnamefont{E.~J.} \bibnamefont{Kim}},
  \bibinfo{author}{\bibfnamefont{J.~E.} \bibnamefont{Tomaszewski}},
  \bibinfo{author}{\bibfnamefont{J.~J.} \bibnamefont{de~Pablo}},
  \bibinfo{author}{\bibfnamefont{P.~F.} \bibnamefont{Nealey}},
  \bibinfo{author}{\bibfnamefont{C.~C.} \bibnamefont{White}}, \bibnamefont{and}
  \bibinfo{author}{\bibfnamefont{W.~L.} \bibnamefont{Wu}},
  \bibinfo{journal}{Macromolecules} \textbf{\bibinfo{volume}{34}},
  \bibinfo{pages}{5627} (\bibinfo{year}{2001}), ISSN \bibinfo{issn}{0024-9297}.

\bibitem[{\citenamefont{Ellison and Torkelson}(2003)}]{et03}
\bibinfo{author}{\bibfnamefont{C.~J.} \bibnamefont{Ellison}} \bibnamefont{and}
  \bibinfo{author}{\bibfnamefont{J.~M.} \bibnamefont{Torkelson}},
  \bibinfo{journal}{Nature Materials} \textbf{\bibinfo{volume}{2}},
  \bibinfo{pages}{695} (\bibinfo{year}{2003}), ISSN \bibinfo{issn}{1476-1122}.

\bibitem[{\citenamefont{Bansal et~al.}(2005)\citenamefont{Bansal, Yang, Li,
  Cho, Benicewicz, Kumar, and Schadler}}]{kumar05}
\bibinfo{author}{\bibfnamefont{A.}~\bibnamefont{Bansal}},
  \bibinfo{author}{\bibfnamefont{H.~C.} \bibnamefont{Yang}},
  \bibinfo{author}{\bibfnamefont{C.~Z.} \bibnamefont{Li}},
  \bibinfo{author}{\bibfnamefont{K.~W.} \bibnamefont{Cho}},
  \bibinfo{author}{\bibfnamefont{B.~C.} \bibnamefont{Benicewicz}},
  \bibinfo{author}{\bibfnamefont{S.~K.} \bibnamefont{Kumar}}, \bibnamefont{and}
  \bibinfo{author}{\bibfnamefont{L.~S.} \bibnamefont{Schadler}},
  \bibinfo{journal}{Nature Materials} \textbf{\bibinfo{volume}{4}},
  \bibinfo{pages}{693} (\bibinfo{year}{2005}), ISSN \bibinfo{issn}{1476-1122}.

\bibitem[{\citenamefont{Priestley et~al.}(2005)\citenamefont{Priestley,
  Ellison, Broadbelt, and Torkelson}}]{tork05}
\bibinfo{author}{\bibfnamefont{R.~D.} \bibnamefont{Priestley}},
  \bibinfo{author}{\bibfnamefont{C.~J.} \bibnamefont{Ellison}},
  \bibinfo{author}{\bibfnamefont{L.~J.} \bibnamefont{Broadbelt}},
  \bibnamefont{and} \bibinfo{author}{\bibfnamefont{J.~M.}
  \bibnamefont{Torkelson}}, \bibinfo{journal}{Science}
  \textbf{\bibinfo{volume}{309}}, \bibinfo{pages}{456} (\bibinfo{year}{2005}),
  ISSN \bibinfo{issn}{0036-8075}.

\bibitem[{\citenamefont{Kim et~al.}(2009)\citenamefont{Kim, Hewlett, Roth, and
  Torkelson}}]{tork09}
\bibinfo{author}{\bibfnamefont{S.}~\bibnamefont{Kim}},
  \bibinfo{author}{\bibfnamefont{S.~A.} \bibnamefont{Hewlett}},
  \bibinfo{author}{\bibfnamefont{C.~B.} \bibnamefont{Roth}}, \bibnamefont{and}
  \bibinfo{author}{\bibfnamefont{J.~M.} \bibnamefont{Torkelson}},
  \bibinfo{journal}{Eur. Phys. J. E} \textbf{\bibinfo{volume}{30}},
  \bibinfo{pages}{83} (\bibinfo{year}{2009}), ISSN \bibinfo{issn}{1292-8941}.

\bibitem[{\citenamefont{Torres et~al.}(2000)\citenamefont{Torres, Nealey, and
  de~Pablo}}]{depablo00-2}
\bibinfo{author}{\bibfnamefont{J.~A.} \bibnamefont{Torres}},
  \bibinfo{author}{\bibfnamefont{P.~F.} \bibnamefont{Nealey}},
  \bibnamefont{and} \bibinfo{author}{\bibfnamefont{J.~J.}
  \bibnamefont{de~Pablo}}, \bibinfo{journal}{Phys. Rev. Lett.}
  \textbf{\bibinfo{volume}{85}}, \bibinfo{pages}{3221} (\bibinfo{year}{2000}),
  ISSN \bibinfo{issn}{0031-9007}.

\bibitem[{\citenamefont{Smith et~al.}(2003)\citenamefont{Smith, Bedrov, and
  Borodin}}]{smith03}
\bibinfo{author}{\bibfnamefont{G.~D.} \bibnamefont{Smith}},
  \bibinfo{author}{\bibfnamefont{D.}~\bibnamefont{Bedrov}}, \bibnamefont{and}
  \bibinfo{author}{\bibfnamefont{O.}~\bibnamefont{Borodin}},
  \bibinfo{journal}{Phys. Rev. Lett.} \textbf{\bibinfo{volume}{90}},
  \bibinfo{pages}{226103} (\bibinfo{year}{2003}),
  \urlprefix\url{http://link.aps.org/doi/10.1103/PhysRevLett.90.226103}.

\bibitem[{\citenamefont{Baschnagel and Varnick}(2005)}]{basch05}
\bibinfo{author}{\bibfnamefont{J.}~\bibnamefont{Baschnagel}} \bibnamefont{and}
  \bibinfo{author}{\bibfnamefont{F.}~\bibnamefont{Varnick}},
  \bibinfo{journal}{Condensed Matter} \textbf{\bibinfo{volume}{17}},
  \bibinfo{pages}{852} (\bibinfo{year}{2005}).

\bibitem[{\citenamefont{Riggleman et~al.}(2006)\citenamefont{Riggleman,
  Yoshimoto, Douglas, and de~Pablo}}]{rigg-prl06}
\bibinfo{author}{\bibfnamefont{R.~A.} \bibnamefont{Riggleman}},
  \bibinfo{author}{\bibfnamefont{K.}~\bibnamefont{Yoshimoto}},
  \bibinfo{author}{\bibfnamefont{J.~F.} \bibnamefont{Douglas}},
  \bibnamefont{and} \bibinfo{author}{\bibfnamefont{J.~J.}
  \bibnamefont{de~Pablo}}, \bibinfo{journal}{Phys. Rev. Lett.}
  \textbf{\bibinfo{volume}{97}}, \bibinfo{pages}{045502}
  (\bibinfo{year}{2006}), ISSN \bibinfo{issn}{0031-9007}.

\bibitem[{\citenamefont{Peter et~al.}(2009)\citenamefont{Peter, Meyer, and
  Baschnagel}}]{basch09}
\bibinfo{author}{\bibfnamefont{S.}~\bibnamefont{Peter}},
  \bibinfo{author}{\bibfnamefont{H.}~\bibnamefont{Meyer}}, \bibnamefont{and}
  \bibinfo{author}{\bibfnamefont{J.}~\bibnamefont{Baschnagel}},
  \bibinfo{journal}{J. Chem. Phys.} \textbf{\bibinfo{volume}{131}},
  \bibinfo{pages}{014902} (\bibinfo{year}{2009}), ISSN
  \bibinfo{issn}{0021-9606}.

\bibitem[{\citenamefont{Barrat et~al.}(2010)\citenamefont{Barrat, Baschnagel,
  and Lyulin}}]{basch-soft-rev10}
\bibinfo{author}{\bibfnamefont{J.-L.} \bibnamefont{Barrat}},
  \bibinfo{author}{\bibfnamefont{J.}~\bibnamefont{Baschnagel}},
  \bibnamefont{and} \bibinfo{author}{\bibfnamefont{A.}~\bibnamefont{Lyulin}},
  \bibinfo{journal}{Soft Matter} \textbf{\bibinfo{volume}{6}},
  \bibinfo{pages}{3430} (\bibinfo{year}{2010}).

\bibitem[{\citenamefont{Forrest et~al.}(1996)\citenamefont{Forrest,
  Dalnoki-Veress, Stevens, and Dutcher}}]{forrest96}
\bibinfo{author}{\bibfnamefont{J.~A.} \bibnamefont{Forrest}},
  \bibinfo{author}{\bibfnamefont{K.}~\bibnamefont{Dalnoki-Veress}},
  \bibinfo{author}{\bibfnamefont{J.~R.} \bibnamefont{Stevens}},
  \bibnamefont{and} \bibinfo{author}{\bibfnamefont{J.~R.}
  \bibnamefont{Dutcher}}, \bibinfo{journal}{Phys. Rev. Lett.}
  \textbf{\bibinfo{volume}{77}}, \bibinfo{pages}{2002} (\bibinfo{year}{1996}),
  ISSN \bibinfo{issn}{0031-9007}.

\bibitem[{\citenamefont{Paeng et~al.}(2012)\citenamefont{Paeng, Richert, and
  Ediger}}]{paeng12}
\bibinfo{author}{\bibfnamefont{K.}~\bibnamefont{Paeng}},
  \bibinfo{author}{\bibfnamefont{R.}~\bibnamefont{Richert}}, \bibnamefont{and}
  \bibinfo{author}{\bibfnamefont{M.~D.} \bibnamefont{Ediger}},
  \bibinfo{journal}{Soft Matter} \textbf{\bibinfo{volume}{8}},
  \bibinfo{pages}{819} (\bibinfo{year}{2012}),
  \urlprefix\url{http://dx.doi.org/10.1039/C1SM06501G}.

\bibitem[{\citenamefont{Ellison et~al.}(2002)\citenamefont{Ellison, Kim, Hall,
  and Torkelson}}]{tork02-2}
\bibinfo{author}{\bibfnamefont{C.~J.} \bibnamefont{Ellison}},
  \bibinfo{author}{\bibfnamefont{S.~D.} \bibnamefont{Kim}},
  \bibinfo{author}{\bibfnamefont{D.~B.} \bibnamefont{Hall}}, \bibnamefont{and}
  \bibinfo{author}{\bibfnamefont{J.~M.} \bibnamefont{Torkelson}},
  \bibinfo{journal}{Eur. Phys. J. E} \textbf{\bibinfo{volume}{8}},
  \bibinfo{pages}{155} (\bibinfo{year}{2002}), ISSN \bibinfo{issn}{1292-8941}.

\bibitem[{\citenamefont{Sch\"{o}nhals et~al.}(2003)\citenamefont{Sch\"{o}nhals,
  Goering, Schick, Frick, and Zorn}}]{schon03}
\bibinfo{author}{\bibfnamefont{A.}~\bibnamefont{Sch\"{o}nhals}},
  \bibinfo{author}{\bibfnamefont{H.}~\bibnamefont{Goering}},
  \bibinfo{author}{\bibfnamefont{C.}~\bibnamefont{Schick}},
  \bibinfo{author}{\bibfnamefont{B.}~\bibnamefont{Frick}}, \bibnamefont{and}
  \bibinfo{author}{\bibfnamefont{R.}~\bibnamefont{Zorn}},
  \bibinfo{journal}{Eur. Phys. J. E} \textbf{\bibinfo{volume}{12}},
  \bibinfo{pages}{173} (\bibinfo{year}{2003}), ISSN \bibinfo{issn}{1292-8941}.

\bibitem[{\citenamefont{Kawana and Jones}(2001)}]{kj01}
\bibinfo{author}{\bibfnamefont{S.}~\bibnamefont{Kawana}} \bibnamefont{and}
  \bibinfo{author}{\bibfnamefont{R.~A.~L.} \bibnamefont{Jones}},
  \bibinfo{journal}{Phys. Rev. E} \textbf{\bibinfo{volume}{63}},
  \bibinfo{pages}{021501} (\bibinfo{year}{2001}), ISSN
  \bibinfo{issn}{1063-651X}.

\bibitem[{\citenamefont{Peter et~al.}(2006)\citenamefont{Peter, Meyer, and
  Baschnagel}}]{basch06}
\bibinfo{author}{\bibfnamefont{S.}~\bibnamefont{Peter}},
  \bibinfo{author}{\bibfnamefont{H.}~\bibnamefont{Meyer}}, \bibnamefont{and}
  \bibinfo{author}{\bibfnamefont{J.}~\bibnamefont{Baschnagel}},
  \bibinfo{journal}{J. Polym. Sci. B} \textbf{\bibinfo{volume}{44}},
  \bibinfo{pages}{2951} (\bibinfo{year}{2006}), ISSN \bibinfo{issn}{0887-6266}.

\bibitem[{\citenamefont{Hanakata et~al.}(2012)\citenamefont{Hanakata, Douglas,
  and Starr}}]{hanakata12}
\bibinfo{author}{\bibfnamefont{P.~Z.} \bibnamefont{Hanakata}},
  \bibinfo{author}{\bibfnamefont{J.~F.} \bibnamefont{Douglas}},
  \bibnamefont{and} \bibinfo{author}{\bibfnamefont{F.~W.} \bibnamefont{Starr}},
  \bibinfo{journal}{The Journal of Chemical Physics}
  \textbf{\bibinfo{volume}{137}}, \bibinfo{eid}{244901} (\bibinfo{year}{2012}),
  \urlprefix\url{http://scitation.aip.org/content/aip/journal/jcp/137/24/10.1063/1.4772402}.

\bibitem[{\citenamefont{Roth et~al.}(2007)\citenamefont{Roth, McNerny, Jager,
  and Torkelson}}]{tork07}
\bibinfo{author}{\bibfnamefont{C.~B.} \bibnamefont{Roth}},
  \bibinfo{author}{\bibfnamefont{K.~L.} \bibnamefont{McNerny}},
  \bibinfo{author}{\bibfnamefont{W.~F.} \bibnamefont{Jager}}, \bibnamefont{and}
  \bibinfo{author}{\bibfnamefont{J.~M.} \bibnamefont{Torkelson}},
  \bibinfo{journal}{Macromolecules} \textbf{\bibinfo{volume}{40}},
  \bibinfo{pages}{2568} (\bibinfo{year}{2007}),
  \eprint{http://pubs.acs.org/doi/pdf/10.1021/ma062864w},
  \urlprefix\url{http://pubs.acs.org/doi/abs/10.1021/ma062864w}.

\bibitem[{\citenamefont{Tate et~al.}(2001)\citenamefont{Tate, Fryer,
  Pasqualini, Montague, de~Pablo, and Nealey}}]{tate01}
\bibinfo{author}{\bibfnamefont{R.~S.} \bibnamefont{Tate}},
  \bibinfo{author}{\bibfnamefont{D.~S.} \bibnamefont{Fryer}},
  \bibinfo{author}{\bibfnamefont{S.}~\bibnamefont{Pasqualini}},
  \bibinfo{author}{\bibfnamefont{M.~F.} \bibnamefont{Montague}},
  \bibinfo{author}{\bibfnamefont{J.~J.} \bibnamefont{de~Pablo}},
  \bibnamefont{and} \bibinfo{author}{\bibfnamefont{P.~F.}
  \bibnamefont{Nealey}}, \bibinfo{journal}{J. Chem. Phys.}
  \textbf{\bibinfo{volume}{115}}, \bibinfo{pages}{9982} (\bibinfo{year}{2001}),
  \urlprefix\url{http://scitation.aip.org/content/aip/journal/jcp/115/21/10.1063/1.1415497}.

\bibitem[{\citenamefont{Erber et~al.}(2010)\citenamefont{Erber, Tress, Mapesa,
  Serghei, Eichhorn, Voit, and Kremer}}]{erber10}
\bibinfo{author}{\bibfnamefont{M.}~\bibnamefont{Erber}},
  \bibinfo{author}{\bibfnamefont{M.}~\bibnamefont{Tress}},
  \bibinfo{author}{\bibfnamefont{E.~U.} \bibnamefont{Mapesa}},
  \bibinfo{author}{\bibfnamefont{A.}~\bibnamefont{Serghei}},
  \bibinfo{author}{\bibfnamefont{K.-J.} \bibnamefont{Eichhorn}},
  \bibinfo{author}{\bibfnamefont{B.}~\bibnamefont{Voit}}, \bibnamefont{and}
  \bibinfo{author}{\bibfnamefont{F.}~\bibnamefont{Kremer}},
  \bibinfo{journal}{Macromolecules} \textbf{\bibinfo{volume}{43}},
  \bibinfo{pages}{7729} (\bibinfo{year}{2010}),
  \eprint{http://pubs.acs.org/doi/pdf/10.1021/ma100912r},
  \urlprefix\url{http://pubs.acs.org/doi/abs/10.1021/ma100912r}.

\bibitem[{\citenamefont{Tong and Sibener}(2013)}]{tong13}
\bibinfo{author}{\bibfnamefont{Q.}~\bibnamefont{Tong}} \bibnamefont{and}
  \bibinfo{author}{\bibfnamefont{S.~J.} \bibnamefont{Sibener}},
  \bibinfo{journal}{Macromolecules} \textbf{\bibinfo{volume}{46}},
  \bibinfo{pages}{8538} (\bibinfo{year}{2013}),
  \eprint{http://pubs.acs.org/doi/pdf/10.1021/ma401629s},
  \urlprefix\url{http://pubs.acs.org/doi/abs/10.1021/ma401629s}.

\bibitem[{\citenamefont{Hudzinskyy et~al.}(2011)\citenamefont{Hudzinskyy,
  Lyulin, Baljon, Balabaev, and Michels}}]{michels11}
\bibinfo{author}{\bibfnamefont{D.}~\bibnamefont{Hudzinskyy}},
  \bibinfo{author}{\bibfnamefont{A.~V.} \bibnamefont{Lyulin}},
  \bibinfo{author}{\bibfnamefont{A.~R.~C.} \bibnamefont{Baljon}},
  \bibinfo{author}{\bibfnamefont{N.~K.} \bibnamefont{Balabaev}},
  \bibnamefont{and} \bibinfo{author}{\bibfnamefont{M.~A.~J.}
  \bibnamefont{Michels}}, \bibinfo{journal}{Macromolecules}
  \textbf{\bibinfo{volume}{44}}, \bibinfo{pages}{2299} (\bibinfo{year}{2011}),
  \eprint{http://pubs.acs.org/doi/pdf/10.1021/ma102567s},
  \urlprefix\url{http://pubs.acs.org/doi/abs/10.1021/ma102567s}.

\bibitem[{\citenamefont{Batistakis et~al.}(2012)\citenamefont{Batistakis,
  Lyulin, and Michels}}]{michels12}
\bibinfo{author}{\bibfnamefont{C.}~\bibnamefont{Batistakis}},
  \bibinfo{author}{\bibfnamefont{A.~V.} \bibnamefont{Lyulin}},
  \bibnamefont{and} \bibinfo{author}{\bibfnamefont{M.~A.~J.}
  \bibnamefont{Michels}}, \bibinfo{journal}{Macromolecules}
  \textbf{\bibinfo{volume}{45}}, \bibinfo{pages}{7282} (\bibinfo{year}{2012}),
  \eprint{http://pubs.acs.org/doi/pdf/10.1021/ma300753e},
  \urlprefix\url{http://pubs.acs.org/doi/abs/10.1021/ma300753e}.

\bibitem[{\citenamefont{Sengupta et~al.}(2013)\citenamefont{Sengupta, Karmakar,
  Dasgupta, and Sastry}}]{sastry13}
\bibinfo{author}{\bibfnamefont{S.}~\bibnamefont{Sengupta}},
  \bibinfo{author}{\bibfnamefont{S.}~\bibnamefont{Karmakar}},
  \bibinfo{author}{\bibfnamefont{C.}~\bibnamefont{Dasgupta}}, \bibnamefont{and}
  \bibinfo{author}{\bibfnamefont{S.}~\bibnamefont{Sastry}},
  \bibinfo{journal}{J. Chem. Phys.} \textbf{\bibinfo{volume}{138}},
  \bibinfo{pages}{12A548} (\bibinfo{year}{2013}).

\bibitem[{\citenamefont{Ediger et~al.}(2008)\citenamefont{Ediger, Harrowell,
  and Yu}}]{ediger08Decoupling}
\bibinfo{author}{\bibfnamefont{M.~D.} \bibnamefont{Ediger}},
  \bibinfo{author}{\bibfnamefont{P.}~\bibnamefont{Harrowell}},
  \bibnamefont{and} \bibinfo{author}{\bibfnamefont{L.}~\bibnamefont{Yu}},
  \bibinfo{journal}{J. Chem. Phys.} \textbf{\bibinfo{volume}{128}},
  \bibinfo{eid}{034709} (\bibinfo{year}{2008}),
  \urlprefix\url{http://scitation.aip.org/content/aip/journal/jcp/128/3/10.1063/1.2815325}.

\bibitem[{\citenamefont{Starr et~al.}(2013)\citenamefont{Starr, Douglas, and
  Sastry}}]{sds13}
\bibinfo{author}{\bibfnamefont{F.~W.} \bibnamefont{Starr}},
  \bibinfo{author}{\bibfnamefont{J.~F.} \bibnamefont{Douglas}},
  \bibnamefont{and} \bibinfo{author}{\bibfnamefont{S.}~\bibnamefont{Sastry}},
  \bibinfo{journal}{J. Chem. Phys.} \textbf{\bibinfo{volume}{138}},
  \bibinfo{pages}{12A541} (\bibinfo{year}{2013}).

\bibitem[{\citenamefont{Starr and Douglas}(2011)}]{sd11}
\bibinfo{author}{\bibfnamefont{F.~W.} \bibnamefont{Starr}} \bibnamefont{and}
  \bibinfo{author}{\bibfnamefont{J.~F.} \bibnamefont{Douglas}},
  \bibinfo{journal}{Phys. Rev. Lett.} \textbf{\bibinfo{volume}{106}},
  \bibinfo{pages}{115702} (\bibinfo{year}{2011}).

\bibitem[{\citenamefont{Pazmino~Betancourt
  et~al.}(2013)\citenamefont{Pazmino~Betancourt, Douglas, and
  Starr}}]{beatriz13}
\bibinfo{author}{\bibfnamefont{B.~A.} \bibnamefont{Pazmino~Betancourt}},
  \bibinfo{author}{\bibfnamefont{J.~F.} \bibnamefont{Douglas}},
  \bibnamefont{and} \bibinfo{author}{\bibfnamefont{F.~W.} \bibnamefont{Starr}},
  \bibinfo{journal}{Soft Matter} \textbf{\bibinfo{volume}{9}},
  \bibinfo{pages}{241} (\bibinfo{year}{2013}),
  \urlprefix\url{http://dx.doi.org/10.1039/C2SM26800K}.

\bibitem[{\citenamefont{Scheidler et~al.}(2002)\citenamefont{Scheidler, Kob,
  and Binder}}]{scheidler02}
\bibinfo{author}{\bibfnamefont{P.}~\bibnamefont{Scheidler}},
  \bibinfo{author}{\bibfnamefont{W.}~\bibnamefont{Kob}}, \bibnamefont{and}
  \bibinfo{author}{\bibfnamefont{K.}~\bibnamefont{Binder}},
  \bibinfo{journal}{Europhys. Lett.} \textbf{\bibinfo{volume}{59}},
  \bibinfo{pages}{701} (\bibinfo{year}{2002}),
  \urlprefix\url{http://stacks.iop.org/0295-5075/59/i=5/a=701}.

\bibitem[{\citenamefont{Virgiliis et~al.}(2012)\citenamefont{Virgiliis,
  Milchev, Rostiashvili, and Vilgis}}]{virgiliis12}
\bibinfo{author}{\bibfnamefont{A.}~\bibnamefont{Virgiliis}},
  \bibinfo{author}{\bibfnamefont{A.}~\bibnamefont{Milchev}},
  \bibinfo{author}{\bibfnamefont{V.}~\bibnamefont{Rostiashvili}},
  \bibnamefont{and} \bibinfo{author}{\bibfnamefont{T.}~\bibnamefont{Vilgis}},
  \bibinfo{journal}{Eur. Phys. J. E} \textbf{\bibinfo{volume}{35}},
  \bibinfo{pages}{1} (\bibinfo{year}{2012}), ISSN \bibinfo{issn}{1292-8941},
  \urlprefix\url{http://dx.doi.org/10.1140/epje/i2012-12097-6}.

\bibitem[{\citenamefont{Angell}(1997)}]{Angell97}
\bibinfo{author}{\bibfnamefont{C.}~\bibnamefont{Angell}},
  \bibinfo{journal}{Polymer} \textbf{\bibinfo{volume}{38}},
  \bibinfo{pages}{6261 } (\bibinfo{year}{1997}), ISSN
  \bibinfo{issn}{0032-3861},
  \urlprefix\url{http://www.sciencedirect.com/science/article/pii/S0032386197002012}.

\bibitem[{\citenamefont{Angell}(1995)}]{angell95}
\bibinfo{author}{\bibfnamefont{C.~A.} \bibnamefont{Angell}},
  \bibinfo{journal}{Science} \textbf{\bibinfo{volume}{267}},
  \bibinfo{pages}{1924} (\bibinfo{year}{1995}).

\bibitem[{\citenamefont{Qin and McKenna}(2006)}]{mckenna}
\bibinfo{author}{\bibfnamefont{Q.}~\bibnamefont{Qin}} \bibnamefont{and}
  \bibinfo{author}{\bibfnamefont{G.~B.} \bibnamefont{McKenna}},
  \bibinfo{journal}{J. Non-Cryst. Solids} \textbf{\bibinfo{volume}{352}},
  \bibinfo{pages}{2977} (\bibinfo{year}{2006}), ISSN \bibinfo{issn}{0022-3093}.

\bibitem[{\citenamefont{Binder et~al.}(2003)\citenamefont{Binder, Baschnagel,
  and Paul}}]{basch03}
\bibinfo{author}{\bibfnamefont{K.}~\bibnamefont{Binder}},
  \bibinfo{author}{\bibfnamefont{J.}~\bibnamefont{Baschnagel}},
  \bibnamefont{and} \bibinfo{author}{\bibfnamefont{W.}~\bibnamefont{Paul}},
  \bibinfo{journal}{Prog. Polym. Sci.} \textbf{\bibinfo{volume}{28}},
  \bibinfo{pages}{115} (\bibinfo{year}{2003}), ISSN \bibinfo{issn}{0079-6700}.

\bibitem[{\citenamefont{Yamakawa}(1971)}]{Yamakawabook}
\bibinfo{author}{\bibfnamefont{H.}~\bibnamefont{Yamakawa}},
  \emph{\bibinfo{title}{Modern Theory of Polymer Solutions}}
  (\bibinfo{publisher}{Harper and Row Publishers}, \bibinfo{year}{1971}).

\bibitem[{\citenamefont{Freed}(1985)}]{freed85}
\bibinfo{author}{\bibfnamefont{K.~F.} \bibnamefont{Freed}},
  \bibinfo{journal}{Journal of Physics A: Mathematical and General}
  \textbf{\bibinfo{volume}{18}}, \bibinfo{pages}{871} (\bibinfo{year}{1985}),
  \urlprefix\url{http://stacks.iop.org/0305-4470/18/i=5/a=019}.

\bibitem[{\citenamefont{Douglas}(1989)}]{douglas89}
\bibinfo{author}{\bibfnamefont{J.~F.} \bibnamefont{Douglas}},
  \bibinfo{journal}{Macromolecules} \textbf{\bibinfo{volume}{22}},
  \bibinfo{pages}{3707} (\bibinfo{year}{1989}),
  \eprint{http://dx.doi.org/10.1021/ma00199a035},
  \urlprefix\url{http://dx.doi.org/10.1021/ma00199a035}.

\bibitem[{\citenamefont{Pazmi\~no Betancourt
  et~al.}(2014{\natexlab{a}})\citenamefont{Pazmi\~no Betancourt, Starr, and
  Douglas}}]{bea-new}
\bibinfo{author}{\bibfnamefont{B.~A.} \bibnamefont{Pazmi\~no Betancourt}},
  \bibinfo{author}{\bibfnamefont{F.~W.} \bibnamefont{Starr}}, \bibnamefont{and}
  \bibinfo{author}{\bibfnamefont{J.~F.} \bibnamefont{Douglas}}
  (\bibinfo{year}{2014}{\natexlab{a}}), \bibinfo{note}{in preparation}.

\bibitem[{\citenamefont{Sokolov and Schweizer}(2009)}]{sokolov09Decoupling}
\bibinfo{author}{\bibfnamefont{A.~P.} \bibnamefont{Sokolov}} \bibnamefont{and}
  \bibinfo{author}{\bibfnamefont{K.~S.} \bibnamefont{Schweizer}},
  \bibinfo{journal}{Phys. Rev. Lett.} \textbf{\bibinfo{volume}{102}},
  \bibinfo{pages}{248301} (\bibinfo{year}{2009}),
  \urlprefix\url{http://link.aps.org/doi/10.1103/PhysRevLett.102.248301}.

\bibitem[{\citenamefont{Urakawa et~al.}(2004)\citenamefont{Urakawa, Swallen,
  Ediger, and von Meerwall}}]{Urakawa04}
\bibinfo{author}{\bibfnamefont{O.}~\bibnamefont{Urakawa}},
  \bibinfo{author}{\bibfnamefont{S.~F.} \bibnamefont{Swallen}},
  \bibinfo{author}{\bibfnamefont{M.~D.} \bibnamefont{Ediger}},
  \bibnamefont{and} \bibinfo{author}{\bibfnamefont{E.~D.} \bibnamefont{von
  Meerwall}}, \bibinfo{journal}{Macromolecules} \textbf{\bibinfo{volume}{37}},
  \bibinfo{pages}{1558} (\bibinfo{year}{2004}),
  \eprint{http://dx.doi.org/10.1021/ma0352025},
  \urlprefix\url{http://dx.doi.org/10.1021/ma0352025}.

\bibitem[{\citenamefont{Mapes et~al.}(2006)\citenamefont{Mapes, Swallen, and
  Ediger}}]{Ediger06}
\bibinfo{author}{\bibfnamefont{M.~K.} \bibnamefont{Mapes}},
  \bibinfo{author}{\bibfnamefont{S.~F.} \bibnamefont{Swallen}},
  \bibnamefont{and} \bibinfo{author}{\bibfnamefont{M.~D.}
  \bibnamefont{Ediger}}, \bibinfo{journal}{The Journal of Physical Chemistry B}
  \textbf{\bibinfo{volume}{110}}, \bibinfo{pages}{507} (\bibinfo{year}{2006}),
  \bibinfo{note}{pMID: 16471562}, \eprint{http://dx.doi.org/10.1021/jp0555955},
  \urlprefix\url{http://dx.doi.org/10.1021/jp0555955}.

\bibitem[{\citenamefont{Douglas and Ishinabe}(1995)}]{sausage}
\bibinfo{author}{\bibfnamefont{J.~F.} \bibnamefont{Douglas}} \bibnamefont{and}
  \bibinfo{author}{\bibfnamefont{T.}~\bibnamefont{Ishinabe}},
  \bibinfo{journal}{Phys. Rev. E} \textbf{\bibinfo{volume}{51}},
  \bibinfo{pages}{1791} (\bibinfo{year}{1995}),
  \urlprefix\url{http://link.aps.org/doi/10.1103/PhysRevE.51.1791}.

\bibitem[{\citenamefont{Adam and Gibbs}(1965)}]{ag}
\bibinfo{author}{\bibfnamefont{G.}~\bibnamefont{Adam}} \bibnamefont{and}
  \bibinfo{author}{\bibfnamefont{J.~H.} \bibnamefont{Gibbs}},
  \bibinfo{journal}{J. Chem. Phys.} \textbf{\bibinfo{volume}{43}},
  \bibinfo{pages}{139} (\bibinfo{year}{1965}), ISSN \bibinfo{issn}{0021-9606}.

\bibitem[{\citenamefont{Pazmi\~no Betancourt
  et~al.}(2014{\natexlab{b}})\citenamefont{Pazmi\~no Betancourt, Douglas, and
  Starr}}]{pds14}
\bibinfo{author}{\bibfnamefont{B.~A.} \bibnamefont{Pazmi\~no Betancourt}},
  \bibinfo{author}{\bibfnamefont{J.~F.} \bibnamefont{Douglas}},
  \bibnamefont{and} \bibinfo{author}{\bibfnamefont{F.~W.} \bibnamefont{Starr}},
  \bibinfo{journal}{The Journal of Chemical Physics}
  \textbf{\bibinfo{volume}{140}}, \bibinfo{eid}{204509}
  (\bibinfo{year}{2014}{\natexlab{b}}),
  \urlprefix\url{http://scitation.aip.org/content/aip/journal/jcp/140/20/10.1063/1.4878502}.

\bibitem[{\citenamefont{Hanakata et~al.}(2014)\citenamefont{Hanakata, Douglas,
  and Starr}}]{hds14}
\bibinfo{author}{\bibfnamefont{P.~Z.} \bibnamefont{Hanakata}},
  \bibinfo{author}{\bibfnamefont{J.~F.} \bibnamefont{Douglas}},
  \bibnamefont{and} \bibinfo{author}{\bibfnamefont{F.~W.} \bibnamefont{Starr}},
  \bibinfo{journal}{Nature Communications} \textbf{\bibinfo{volume}{5}},
  \bibinfo{pages}{256207} (\bibinfo{year}{2014}).

\bibitem[{not()}]{note1}
\bibinfo{note}{In their original work, AG identified $z^*$ with the minimal
  scale of cooperative movement while we identify $z$ with the average over a
  highly polydisperse distribution of collective movements observed in our
  simulations. This is an important conceptual difference between the string
  model and the AG model, but both models lead to Eq. ~\ref{eq:ag} as their
  final result so that the string model is certainly preserves the spirit of
  the AG relation.}

\bibitem[{\citenamefont{Freed}(2014)}]{Freed14}
\bibinfo{author}{\bibfnamefont{K.~F.} \bibnamefont{Freed}},
  \bibinfo{journal}{The Journal of Chemical Physics}
  \textbf{\bibinfo{volume}{141}}, \bibinfo{eid}{141102} (\bibinfo{year}{2014}),
  \urlprefix\url{http://scitation.aip.org/content/aip/journal/jcp/141/14/10.1063/1.4897973}.

\bibitem[{\citenamefont{Donati et~al.}(1998)\citenamefont{Donati, Douglas, Kob,
  Plimpton, Poole, and Glotzer}}]{donati98}
\bibinfo{author}{\bibfnamefont{C.}~\bibnamefont{Donati}},
  \bibinfo{author}{\bibfnamefont{J.~F.} \bibnamefont{Douglas}},
  \bibinfo{author}{\bibfnamefont{W.}~\bibnamefont{Kob}},
  \bibinfo{author}{\bibfnamefont{S.~J.} \bibnamefont{Plimpton}},
  \bibinfo{author}{\bibfnamefont{P.~H.} \bibnamefont{Poole}}, \bibnamefont{and}
  \bibinfo{author}{\bibfnamefont{S.~C.} \bibnamefont{Glotzer}},
  \bibinfo{journal}{Phys. Rev. Lett.} \textbf{\bibinfo{volume}{80}},
  \bibinfo{pages}{2338} (\bibinfo{year}{1998}),
  \urlprefix\url{http://link.aps.org/doi/10.1103/PhysRevLett.80.2338}.

\bibitem[{\citenamefont{Aichele et~al.}(2003)\citenamefont{Aichele,
  Gebremichael, Starr, Baschnagel, and Glotzer}}]{aichele03}
\bibinfo{author}{\bibfnamefont{M.}~\bibnamefont{Aichele}},
  \bibinfo{author}{\bibfnamefont{Y.}~\bibnamefont{Gebremichael}},
  \bibinfo{author}{\bibfnamefont{F.}~\bibnamefont{Starr}},
  \bibinfo{author}{\bibfnamefont{J.}~\bibnamefont{Baschnagel}},
  \bibnamefont{and} \bibinfo{author}{\bibfnamefont{S.}~\bibnamefont{Glotzer}},
  \bibinfo{journal}{J. Chem. Phys.} \textbf{\bibinfo{volume}{119}},
  \bibinfo{pages}{5290} (\bibinfo{year}{2003}).

\bibitem[{\citenamefont{Starr et~al.}(2014)\citenamefont{Starr, Hanakata,
  Betancourt, Sastry, and Douglas}}]{sdspb14}
\bibinfo{author}{\bibfnamefont{F.~W.} \bibnamefont{Starr}},
  \bibinfo{author}{\bibfnamefont{P.~Z.} \bibnamefont{Hanakata}},
  \bibinfo{author}{\bibfnamefont{B.~A.~P.} \bibnamefont{Betancourt}},
  \bibinfo{author}{\bibfnamefont{S.}~\bibnamefont{Sastry}}, \bibnamefont{and}
  \bibinfo{author}{\bibfnamefont{J.~F.} \bibnamefont{Douglas}},
  \emph{\bibinfo{title}{Fragility and Cooperative Motion in Polymer Glass
  Formation}} (\bibinfo{publisher}{Fragility of glass forming liquids},
  \bibinfo{year}{2014}).

\bibitem[{\citenamefont{Pazmi\~no Betancourt
  et~al.}(2014{\natexlab{c}})\citenamefont{Pazmi\~no Betancourt, Hanakata,
  Starr, and Douglas}}]{pds14pnas}
\bibinfo{author}{\bibfnamefont{B.~A.} \bibnamefont{Pazmi\~no Betancourt}},
  \bibinfo{author}{\bibfnamefont{P.~Z.} \bibnamefont{Hanakata}},
  \bibinfo{author}{\bibfnamefont{F.~W.} \bibnamefont{Starr}}, \bibnamefont{and}
  \bibinfo{author}{\bibfnamefont{J.~F.} \bibnamefont{Douglas}}
  (\bibinfo{year}{2014}{\natexlab{c}}), \bibinfo{note}{submitted}.

\bibitem[{\citenamefont{Eyring}(1936)}]{eyring1936}
\bibinfo{author}{\bibfnamefont{H.}~\bibnamefont{Eyring}}, \bibinfo{journal}{J.
  Chem. Phys.} \textbf{\bibinfo{volume}{4}}, \bibinfo{pages}{283}
  (\bibinfo{year}{1936}),
  \urlprefix\url{http://scitation.aip.org/content/aip/journal/jcp/4/4/10.1063/1.1749836}.

\bibitem[{\citenamefont{Ewell}(1938)}]{ewell1938}
\bibinfo{author}{\bibfnamefont{R.~H.} \bibnamefont{Ewell}},
  \bibinfo{journal}{J. Appl. Phys.} \textbf{\bibinfo{volume}{9}},
  \bibinfo{pages}{252} (\bibinfo{year}{1938}),
  \urlprefix\url{http://scitation.aip.org/content/aip/journal/jap/9/4/10.1063/1.1710415}.

\bibitem[{\citenamefont{Glasstone
  et~al.}(1941{\natexlab{a}})\citenamefont{Glasstone, Laidler, and
  Eyring}}]{eyring1941-RateProc}
\bibinfo{author}{\bibfnamefont{S.}~\bibnamefont{Glasstone}},
  \bibinfo{author}{\bibfnamefont{K.~J.} \bibnamefont{Laidler}},
  \bibnamefont{and} \bibinfo{author}{\bibfnamefont{H.}~\bibnamefont{Eyring}},
  \emph{\bibinfo{title}{{Theory of Rate Processes}}}
  (\bibinfo{publisher}{McGraw Hill}, \bibinfo{year}{1941}{\natexlab{a}}),
  \bibinfo{edition}{first edition} ed.,
  \urlprefix\url{http://www.amazon.com/exec/obidos/redirect?tag=citeulike07-20\&path=ASIN/B000IXRJ6W}.

\bibitem[{\citenamefont{Kauzmann and Eyring}(1940)}]{eyringKuzman1940}
\bibinfo{author}{\bibfnamefont{W.}~\bibnamefont{Kauzmann}} \bibnamefont{and}
  \bibinfo{author}{\bibfnamefont{H.}~\bibnamefont{Eyring}},
  \bibinfo{journal}{J. Am. Chem. Soc.} \textbf{\bibinfo{volume}{62}},
  \bibinfo{pages}{3113} (\bibinfo{year}{1940}),
  \eprint{http://pubs.acs.org/doi/pdf/10.1021/ja01868a059},
  \urlprefix\url{http://pubs.acs.org/doi/abs/10.1021/ja01868a059}.

\bibitem[{\citenamefont{Eyring et~al.}(1958)\citenamefont{Eyring, Ree, and
  Hirai}}]{eyring1958}
\bibinfo{author}{\bibfnamefont{H.}~\bibnamefont{Eyring}},
  \bibinfo{author}{\bibfnamefont{T.}~\bibnamefont{Ree}}, \bibnamefont{and}
  \bibinfo{author}{\bibfnamefont{N.}~\bibnamefont{Hirai}},
  \bibinfo{journal}{Proc. Natl. Acad. Sci. USA} \textbf{\bibinfo{volume}{44}},
  \bibinfo{pages}{1213} (\bibinfo{year}{1958}),
  \eprint{http://www.pnas.org/content/44/12/1213.full.pdf+html},
  \urlprefix\url{http://www.pnas.org/content/44/12/1213.short}.

\bibitem[{\citenamefont{Bondi}(1946)}]{bondi1946}
\bibinfo{author}{\bibfnamefont{A.}~\bibnamefont{Bondi}}, \bibinfo{journal}{J.
  Chem. Phys.} \textbf{\bibinfo{volume}{14}}, \bibinfo{pages}{591}
  (\bibinfo{year}{1946}),
  \urlprefix\url{http://scitation.aip.org/content/aip/journal/jcp/14/10/10.1063/1.1724071}.

\bibitem[{\citenamefont{Yelon et~al.}(2006)\citenamefont{Yelon, Movaghar, and
  Crandall}}]{yelon06}
\bibinfo{author}{\bibfnamefont{A.}~\bibnamefont{Yelon}},
  \bibinfo{author}{\bibfnamefont{B.}~\bibnamefont{Movaghar}}, \bibnamefont{and}
  \bibinfo{author}{\bibfnamefont{R.~S.} \bibnamefont{Crandall}},
  \bibinfo{journal}{Rep. Prog. Phys.} \textbf{\bibinfo{volume}{69}},
  \bibinfo{pages}{1145} (\bibinfo{year}{2006}),
  \urlprefix\url{http://stacks.iop.org/0034-4885/69/i=4/a=R04}.

\bibitem[{\citenamefont{Yelon et~al.}(2011)\citenamefont{Yelon, Sacher, and
  Linert}}]{yelon11}
\bibinfo{author}{\bibfnamefont{A.}~\bibnamefont{Yelon}},
  \bibinfo{author}{\bibfnamefont{E.}~\bibnamefont{Sacher}}, \bibnamefont{and}
  \bibinfo{author}{\bibfnamefont{W.}~\bibnamefont{Linert}},
  \bibinfo{journal}{Catalysis Letters} \textbf{\bibinfo{volume}{141}},
  \bibinfo{pages}{954} (\bibinfo{year}{2011}), ISSN \bibinfo{issn}{1011-372X},
  \urlprefix\url{http://dx.doi.org/10.1007/s10562-011-0645-8}.

\bibitem[{\citenamefont{Yelon and Movaghar}(1990)}]{yelon1990}
\bibinfo{author}{\bibfnamefont{A.}~\bibnamefont{Yelon}} \bibnamefont{and}
  \bibinfo{author}{\bibfnamefont{B.}~\bibnamefont{Movaghar}},
  \bibinfo{journal}{Phys. Rev. Lett.} \textbf{\bibinfo{volume}{65}},
  \bibinfo{pages}{618} (\bibinfo{year}{1990}),
  \urlprefix\url{http://link.aps.org/doi/10.1103/PhysRevLett.65.618}.

\bibitem[{\citenamefont{Madge}(1934{\natexlab{a}})}]{madge1934}
\bibinfo{author}{\bibfnamefont{E.~W.} \bibnamefont{Madge}},
  \bibinfo{journal}{J. Appl. Phys.} \textbf{\bibinfo{volume}{5}},
  \bibinfo{pages}{39} (\bibinfo{year}{1934}{\natexlab{a}}),
  \urlprefix\url{http://scitation.aip.org/content/aip/journal/jap/5/2/10.1063/1.1745228}.

\bibitem[{\citenamefont{Eirich and Simha}(1939{\natexlab{a}})}]{eirich1939}
\bibinfo{author}{\bibfnamefont{F.}~\bibnamefont{Eirich}} \bibnamefont{and}
  \bibinfo{author}{\bibfnamefont{R.}~\bibnamefont{Simha}}, \bibinfo{journal}{J.
  Chem. Phys.} \textbf{\bibinfo{volume}{7}}, \bibinfo{pages}{116}
  (\bibinfo{year}{1939}{\natexlab{a}}),
  \urlprefix\url{http://scitation.aip.org/content/aip/journal/jcp/7/2/10.1063/1.1750389}.

\bibitem[{\citenamefont{Qun-Fang
  et~al.}(1997{\natexlab{a}})\citenamefont{Qun-Fang, Yu-Chun, and
  Rui-Sen}}]{qunfang1997}
\bibinfo{author}{\bibfnamefont{L.}~\bibnamefont{Qun-Fang}},
  \bibinfo{author}{\bibfnamefont{H.}~\bibnamefont{Yu-Chun}}, \bibnamefont{and}
  \bibinfo{author}{\bibfnamefont{L.}~\bibnamefont{Rui-Sen}},
  \bibinfo{journal}{Fluid Phase Equilibria} \textbf{\bibinfo{volume}{140}},
  \bibinfo{pages}{221 } (\bibinfo{year}{1997}{\natexlab{a}}), ISSN
  \bibinfo{issn}{0378-3812},
  \urlprefix\url{http://www.sciencedirect.com/science/article/pii/S0378381297001763}.

\bibitem[{\citenamefont{Madge}(1934{\natexlab{b}})}]{Madge34}
\bibinfo{author}{\bibfnamefont{E.~W.} \bibnamefont{Madge}},
  \bibinfo{journal}{Journal of Applied Physics} \textbf{\bibinfo{volume}{5}},
  \bibinfo{pages}{39} (\bibinfo{year}{1934}{\natexlab{b}}),
  \urlprefix\url{http://scitation.aip.org/content/aip/journal/jap/5/2/10.1063/1.1745228}.

\bibitem[{\citenamefont{Eirich and Simha}(1939{\natexlab{b}})}]{Simha39}
\bibinfo{author}{\bibfnamefont{F.}~\bibnamefont{Eirich}} \bibnamefont{and}
  \bibinfo{author}{\bibfnamefont{R.}~\bibnamefont{Simha}},
  \bibinfo{journal}{The Journal of Chemical Physics}
  \textbf{\bibinfo{volume}{7}}, \bibinfo{pages}{116}
  (\bibinfo{year}{1939}{\natexlab{b}}),
  \urlprefix\url{http://scitation.aip.org/content/aip/journal/jcp/7/2/10.1063/1.1750389}.

\bibitem[{\citenamefont{Glasstone
  et~al.}(1941{\natexlab{b}})\citenamefont{Glasstone, Laidler, and
  Eyring}}]{glasstone1941theory}
\bibinfo{author}{\bibfnamefont{S.}~\bibnamefont{Glasstone}},
  \bibinfo{author}{\bibfnamefont{K.}~\bibnamefont{Laidler}}, \bibnamefont{and}
  \bibinfo{author}{\bibfnamefont{H.}~\bibnamefont{Eyring}},
  \emph{\bibinfo{title}{The Theory of Rate Processes: The Kinetics of Chemical
  Reactions, Viscosity, Diffusion and Electrochemical Phenomena}},
  International chemical series (\bibinfo{publisher}{McGraw-Hill Book Company,
  Incorporated}, \bibinfo{year}{1941}{\natexlab{b}}),
  \urlprefix\url{http://books.google.com/books?id=zb2GAAAAIAAJ}.

\bibitem[{\citenamefont{Iwashita
  et~al.}(2013{\natexlab{a}})\citenamefont{Iwashita, Nicholson, and
  Egami}}]{Egami13}
\bibinfo{author}{\bibfnamefont{T.}~\bibnamefont{Iwashita}},
  \bibinfo{author}{\bibfnamefont{D.~M.} \bibnamefont{Nicholson}},
  \bibnamefont{and} \bibinfo{author}{\bibfnamefont{T.}~\bibnamefont{Egami}},
  \bibinfo{journal}{Phys. Rev. Lett.} \textbf{\bibinfo{volume}{110}},
  \bibinfo{pages}{205504} (\bibinfo{year}{2013}{\natexlab{a}}),
  \urlprefix\url{http://link.aps.org/doi/10.1103/PhysRevLett.110.205504}.

\bibitem[{\citenamefont{Qun-Fang
  et~al.}(1997{\natexlab{b}})\citenamefont{Qun-Fang, Yu-Chun, and
  Rui-Sen}}]{QunFang97}
\bibinfo{author}{\bibfnamefont{L.}~\bibnamefont{Qun-Fang}},
  \bibinfo{author}{\bibfnamefont{H.}~\bibnamefont{Yu-Chun}}, \bibnamefont{and}
  \bibinfo{author}{\bibfnamefont{L.}~\bibnamefont{Rui-Sen}},
  \bibinfo{journal}{Fluid Phase Equilibria} \textbf{\bibinfo{volume}{140}},
  \bibinfo{pages}{221 } (\bibinfo{year}{1997}{\natexlab{b}}), ISSN
  \bibinfo{issn}{0378-3812},
  \urlprefix\url{http://www.sciencedirect.com/science/article/pii/S0378381297001763}.

\bibitem[{\citenamefont{Speedy et~al.}(1989)\citenamefont{Speedy, Prielmeier,
  Vardag, Lang, and L\"{u}demann}}]{speedy1989}
\bibinfo{author}{\bibfnamefont{R.}~\bibnamefont{Speedy}},
  \bibinfo{author}{\bibfnamefont{F.}~\bibnamefont{Prielmeier}},
  \bibinfo{author}{\bibfnamefont{T.}~\bibnamefont{Vardag}},
  \bibinfo{author}{\bibfnamefont{E.}~\bibnamefont{Lang}}, \bibnamefont{and}
  \bibinfo{author}{\bibfnamefont{H.-D.} \bibnamefont{L\"{u}demann}},
  \bibinfo{journal}{Mol. Phys.} \textbf{\bibinfo{volume}{66}},
  \bibinfo{pages}{577} (\bibinfo{year}{1989}),
  \urlprefix\url{http://www.tandfonline.com/doi/abs/10.1080/00268978900100341}.

\bibitem[{\citenamefont{{Hentschel} et~al.}(2012)\citenamefont{{Hentschel},
  {Karmakar}, {Procaccia}, and {Zylberg}}}]{hentschel12}
\bibinfo{author}{\bibfnamefont{H.~G.~E.} \bibnamefont{{Hentschel}}},
  \bibinfo{author}{\bibfnamefont{S.}~\bibnamefont{{Karmakar}}},
  \bibinfo{author}{\bibfnamefont{I.}~\bibnamefont{{Procaccia}}},
  \bibnamefont{and}
  \bibinfo{author}{\bibfnamefont{J.}~\bibnamefont{{Zylberg}}},
  \bibinfo{journal}{ArXiv e-prints}  (\bibinfo{year}{2012}),
  \eprint{1202.1127}.

\bibitem[{\citenamefont{Iwashita
  et~al.}(2013{\natexlab{b}})\citenamefont{Iwashita, Nicholson, and
  Egami}}]{iwashita13}
\bibinfo{author}{\bibfnamefont{T.}~\bibnamefont{Iwashita}},
  \bibinfo{author}{\bibfnamefont{D.~M.} \bibnamefont{Nicholson}},
  \bibnamefont{and} \bibinfo{author}{\bibfnamefont{T.}~\bibnamefont{Egami}},
  \bibinfo{journal}{Phys. Rev. Lett.} \textbf{\bibinfo{volume}{110}},
  \bibinfo{pages}{205504} (\bibinfo{year}{2013}{\natexlab{b}}),
  \urlprefix\url{http://link.aps.org/doi/10.1103/PhysRevLett.110.205504}.

\bibitem[{\citenamefont{Boisvert et~al.}(1998)\citenamefont{Boisvert, Mousseau,
  and Lewis}}]{boisvert1998}
\bibinfo{author}{\bibfnamefont{G.}~\bibnamefont{Boisvert}},
  \bibinfo{author}{\bibfnamefont{N.}~\bibnamefont{Mousseau}}, \bibnamefont{and}
  \bibinfo{author}{\bibfnamefont{L.~J.} \bibnamefont{Lewis}},
  \bibinfo{journal}{Phys. Rev. Lett.} \textbf{\bibinfo{volume}{80}},
  \bibinfo{pages}{203} (\bibinfo{year}{1998}),
  \urlprefix\url{http://link.aps.org/doi/10.1103/PhysRevLett.80.203}.

\bibitem[{\citenamefont{Digilov and Reiner}(2004)}]{digilov04}
\bibinfo{author}{\bibfnamefont{R.~M.} \bibnamefont{Digilov}} \bibnamefont{and}
  \bibinfo{author}{\bibfnamefont{M.}~\bibnamefont{Reiner}},
  \bibinfo{journal}{Eur. J. Phys.} \textbf{\bibinfo{volume}{25}},
  \bibinfo{pages}{15} (\bibinfo{year}{2004}),
  \urlprefix\url{http://stacks.iop.org/0143-0807/25/i=1/a=003}.

\bibitem[{\citenamefont{Nash}(1984)}]{nash1984}
\bibinfo{author}{\bibfnamefont{L.~K.} \bibnamefont{Nash}}, \bibinfo{journal}{J.
  Chem. Ed.} \textbf{\bibinfo{volume}{61}}, \bibinfo{pages}{981}
  (\bibinfo{year}{1984}),
  \eprint{http://pubs.acs.org/doi/pdf/10.1021/ed061p981},
  \urlprefix\url{http://pubs.acs.org/doi/abs/10.1021/ed061p981}.

\bibitem[{\citenamefont{Barclay and Butler}(1938)}]{barclay1938}
\bibinfo{author}{\bibfnamefont{I.~M.} \bibnamefont{Barclay}} \bibnamefont{and}
  \bibinfo{author}{\bibfnamefont{J.~A.~V.} \bibnamefont{Butler}},
  \bibinfo{journal}{Trans. Faraday Soc.} \textbf{\bibinfo{volume}{34}},
  \bibinfo{pages}{1445} (\bibinfo{year}{1938}),
  \urlprefix\url{http://dx.doi.org/10.1039/TF9383401445}.

\bibitem[{\citenamefont{Bell}(1937)}]{bell1937}
\bibinfo{author}{\bibfnamefont{R.~P.} \bibnamefont{Bell}},
  \bibinfo{journal}{Trans. Faraday Soc.} \textbf{\bibinfo{volume}{33}},
  \bibinfo{pages}{496} (\bibinfo{year}{1937}),
  \urlprefix\url{http://dx.doi.org/10.1039/TF9373300496}.

\bibitem[{\citenamefont{Evans and Polanyi}(1936)}]{evans1936}
\bibinfo{author}{\bibfnamefont{M.~G.} \bibnamefont{Evans}} \bibnamefont{and}
  \bibinfo{author}{\bibfnamefont{M.}~\bibnamefont{Polanyi}},
  \bibinfo{journal}{Trans. Faraday Soc.} \textbf{\bibinfo{volume}{32}},
  \bibinfo{pages}{1333} (\bibinfo{year}{1936}),
  \urlprefix\url{http://dx.doi.org/10.1039/TF9363201333}.

\bibitem[{\citenamefont{Frank}(1945)}]{henry1945}
\bibinfo{author}{\bibfnamefont{H.~S.} \bibnamefont{Frank}},
  \bibinfo{journal}{J. Chem. Phys.} \textbf{\bibinfo{volume}{13}},
  \bibinfo{pages}{493} (\bibinfo{year}{1945}),
  \urlprefix\url{http://scitation.aip.org/content/aip/journal/jcp/13/11/10.1063/1.1723984}.

\bibitem[{\citenamefont{Barrer}(1943)}]{barrer1943}
\bibinfo{author}{\bibfnamefont{R.~M.} \bibnamefont{Barrer}},
  \bibinfo{journal}{Trans. Faraday Soc.} \textbf{\bibinfo{volume}{39}},
  \bibinfo{pages}{48} (\bibinfo{year}{1943}),
  \urlprefix\url{http://dx.doi.org/10.1039/TF9433900048}.

\bibitem[{\citenamefont{Waring and Becher}(1947)}]{waring1947}
\bibinfo{author}{\bibfnamefont{C.~E.} \bibnamefont{Waring}} \bibnamefont{and}
  \bibinfo{author}{\bibfnamefont{P.}~\bibnamefont{Becher}},
  \bibinfo{journal}{J. Chem. Phys.} \textbf{\bibinfo{volume}{15}},
  \bibinfo{pages}{488} (\bibinfo{year}{1947}),
  \urlprefix\url{http://scitation.aip.org/content/aip/journal/jcp/15/7/10.1063/1.1746569}.

\bibitem[{\citenamefont{Dyre}(1986)}]{dyre1986}
\bibinfo{author}{\bibfnamefont{J.~C.} \bibnamefont{Dyre}}, \bibinfo{journal}{J.
  Phys. C} \textbf{\bibinfo{volume}{19}}, \bibinfo{pages}{5655}
  (\bibinfo{year}{1986}),
  \urlprefix\url{http://stacks.iop.org/0022-3719/19/i=28/a=016}.

\bibitem[{\citenamefont{Eby}(1962)}]{eby62}
\bibinfo{author}{\bibfnamefont{R.~K.} \bibnamefont{Eby}}, \bibinfo{journal}{The
  Journal of Chemical Physics} \textbf{\bibinfo{volume}{37}},
  \bibinfo{pages}{2785} (\bibinfo{year}{1962}),
  \urlprefix\url{http://scitation.aip.org/content/aip/journal/jcp/37/12/10.1063/1.1733106}.

\bibitem[{\citenamefont{Dienes}(1950)}]{dienes50}
\bibinfo{author}{\bibfnamefont{G.~J.} \bibnamefont{Dienes}},
  \bibinfo{journal}{Journal of Applied Physics} \textbf{\bibinfo{volume}{21}},
  \bibinfo{pages}{1189} (\bibinfo{year}{1950}),
  \urlprefix\url{http://scitation.aip.org/content/aip/journal/jap/21/11/10.1063/1.1699563}.

\bibitem[{\citenamefont{Fisher and Ferdinand}(1967)}]{ff87}
\bibinfo{author}{\bibfnamefont{M.~E.} \bibnamefont{Fisher}} \bibnamefont{and}
  \bibinfo{author}{\bibfnamefont{A.~E.} \bibnamefont{Ferdinand}},
  \bibinfo{journal}{Phys. Rev. Lett.} \textbf{\bibinfo{volume}{19}},
  \bibinfo{pages}{169} (\bibinfo{year}{1967}),
  \urlprefix\url{http://link.aps.org/doi/10.1103/PhysRevLett.19.169}.

\bibitem[{\citenamefont{Allan}(1970)}]{allan70}
\bibinfo{author}{\bibfnamefont{G.~A.~T.} \bibnamefont{Allan}},
  \bibinfo{journal}{Phys. Rev. B} \textbf{\bibinfo{volume}{1}},
  \bibinfo{pages}{352} (\bibinfo{year}{1970}),
  \urlprefix\url{http://link.aps.org/doi/10.1103/PhysRevB.1.352}.

\bibitem[{\citenamefont{Jackson and McKenna}(1990)}]{mackena90}
\bibinfo{author}{\bibfnamefont{C.~L.} \bibnamefont{Jackson}} \bibnamefont{and}
  \bibinfo{author}{\bibfnamefont{G.~B.} \bibnamefont{McKenna}},
  \bibinfo{journal}{The Journal of Chemical Physics}
  \textbf{\bibinfo{volume}{93}}, \bibinfo{pages}{9002} (\bibinfo{year}{1990}),
  \urlprefix\url{http://scitation.aip.org/content/aip/journal/jcp/93/12/10.1063/1.459240}.

\bibitem[{\citenamefont{Zhang and Douglas}(2013)}]{zhang13}
\bibinfo{author}{\bibfnamefont{H.}~\bibnamefont{Zhang}} \bibnamefont{and}
  \bibinfo{author}{\bibfnamefont{J.~F.} \bibnamefont{Douglas}},
  \bibinfo{journal}{Soft Matter} \textbf{\bibinfo{volume}{9}},
  \bibinfo{pages}{1254} (\bibinfo{year}{2013}),
  \urlprefix\url{http://dx.doi.org/10.1039/C2SM26789F}.

\end{thebibliography}
\end{document}